\documentclass[acmsmall]{acmart}
\AtBeginDocument{%
  }

\setcopyright{acmlicensed}
\copyrightyear{2026}
\acmYear{2026}
\acmDOI{XXXXXXX.XXXXXXX}

\acmJournal{JACM}
\acmVolume{37}
\acmNumber{4}
\acmArticle{111}
\acmMonth{7}

\usepackage{graphicx} 
\usepackage{caption}
\usepackage{booktabs}   
\usepackage{tabularx}   
\usepackage{array}      
\usepackage{multirow}   
\usepackage[normalem]{ulem}
\useunder{\uline}{\ul}{}
\usepackage[table,xcdraw]{xcolor}

\usepackage{mdframed}
\usepackage[most]{tcolorbox}
\usepackage{xcolor}
\usepackage{fancyvrb}

\definecolor{titleblue}{RGB}{54,95,145}
\definecolor{titlebg}{RGB}{235,243,252}

\definecolor{riskred}{RGB}{176,62,62}
\definecolor{riskbg}{RGB}{253,239,239}

\definecolor{safegreen}{RGB}{46,132,92}
\definecolor{safebg}{RGB}{238,250,244}

\begin{document}

\title{Security in LLM-Assisted Coding: A Systematic Literature Review}

\author{Liang Yuan}
\authornote{Corresponding author}
\email{liang.yuan@acu.edu.au}
\orcid{0000-0002-6766-7913}

\author{Maryam Khanian Najafabadi}
\email{maryam.khanian@acu.edu.au}
\orcid{0000-0002-5071-7515}

\author{Mashud Rana}
\email{mashud.rana@acu.edu.au}
\orcid{0000-0003-2999-9367}

\author{Varvara Vetrova}
\email{varvara.vetrova@acu.edu.au}
\orcid{0000-0001-8538-5584}

\affiliation{%
  \institution{Australian Catholic University}
  \city{North Sydney}
  \state{New South Wales}
  \country{Australia}
}

\renewcommand{\shortauthors}{Yuan et al.}

\begin{abstract}
  Large Language Models accelerate software development but introduce severe security vulnerabilities. This systematic review investigates security risks in automated coding, benchmark limitations, and the role of human developers. Findings indicate that models frequently generate insecure code and faulty patches due to training data biases. Additionally, the models themselves are susceptible to direct attacks. Current evaluation methods suffer from data contamination and lack robust security metrics. Consequently, human oversight remains essential. Safer integration of automation will require more security-aware model architectures, stronger security-specific benchmarks, and more effective human-AI collaboration frameworks to mitigate these systemic risks.
\end{abstract}


  

\maketitle

\section{Introduction}
The advent of Large Language Models (LLMs) marks a transformative period in software engineering, offering unprecedented capabilities in automating and assisting various coding tasks. From code generation and completion~\cite{kharma2026security, mu2024clarifygpt, jiang2024self, ding2024cocomic} to automated program repair (APR) and vulnerability patching~\cite{pearce2023examining, zhou2024out, fu2024vision, wei2023copiloting}, LLMs promise to enhance developer productivity and accelerate software development lifecycles~\cite{barke2023grounded}. However, this rapid integration is paralleled by a growing array of security concerns that demand systematic investigation and proactive mitigation strategies~\cite{he2023large}.

A central challenge in securing LLM-assisted coding is that functional correctness does not establish software security. Generated code may satisfy its intended functionality while containing exploitable vulnerabilities, and an automatically produced patch may leave the underlying security issue incompletely resolved or introduce additional weaknesses~\cite{pearce2022asleep, gong2024well, liu2024no}. Assessing these risks is further complicated by limitations in evaluation datasets and pipelines, including benchmark contamination, incomplete vulnerability coverage, and differences between isolated coding tasks and realistic development settings. Human review provides an additional safeguard, but its effectiveness depends on developers' security expertise and their ability to critically assess model outputs~\cite{mohsin2024can, perry2023users}.

These challenges involve two related but distinct concepts: \emph{software security} and \emph{AI safety}. Software security concerns the confidentiality, integrity, and availability of software artifacts and systems, including vulnerabilities introduced through code generation, insecure repairs, and adversarial manipulation. AI safety encompasses broader concerns about the dependable, controllable, and socially responsible operation of LLM-based systems, including misuse resistance, human oversight, trust calibration, accountability, and governance~\cite{chow2025large}. The distinction matters because a coding system's security depends both on the properties of its outputs and on the conditions under which those outputs are produced, evaluated, and accepted into software systems.

The transition from passive code-completion tools to \emph{agentic coding systems} makes these interactions more consequential. Autonomous coding agents, multi-agent frameworks, and tool-using LLMs can search repositories, invoke external tools, execute tests, edit files, and iteratively revise their outputs~\cite{chow2026dialogue, liu2024agents4plc, nunez2024autosafecoder, zheng2024opencodeinterpreter}. These capabilities expand the security problem beyond individual code snippets. Errors can propagate across multiple steps, tool invocation and environment access introduce additional attack surfaces, and increasing autonomy can make intermediate actions less visible to human reviewers. Secure deployment therefore requires attention to workflow-level risks, the limits of human oversight, and the controls governing an agent's actions.

\begin{figure}[t]
\centering
    \scalebox{0.9}{
    \includegraphics[width=1\linewidth]{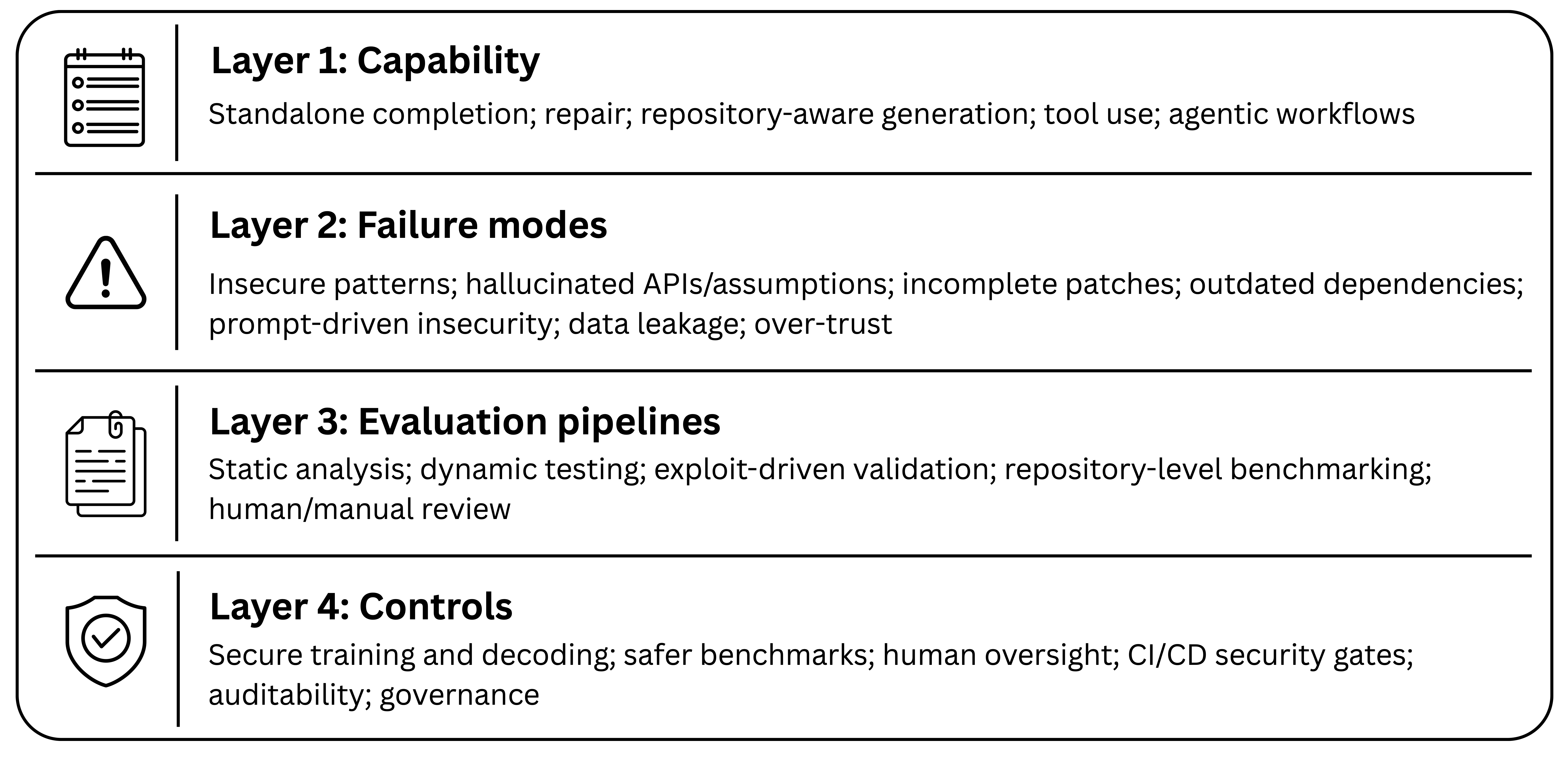}
    }
    \vspace{-0.4cm}
    \caption{A layered framework for analyzing security in LLM-assisted coding.} 
\label{fig:layered-framework}
\vspace{-0.5cm}
\end{figure}

Existing studies provide substantial coverage of individual parts of this problem. Reviews of automated program repair examine repair techniques and their limitations~\cite{huang2023survey, anand2024comprehensive, zhang2024systematic}, while code-security surveys address vulnerabilities in generated code and the use of LLMs for vulnerability detection and repair~\cite{basic2025vulnerabilities, sheng2025llms, zhou2024large}. Broader reviews of AI safety and agentic systems examine oversight, controllability, and governance~\cite{chow2025large, chow2026dialogue}. However, these strands provide limited synthesis of how code-level weaknesses, evaluation limitations, human interaction, and increasing autonomy jointly affect security in coding workflows. In particular, the relationship between what current evaluations measure and what is needed to support secure deployment remains insufficiently integrated with analyses of human oversight and agentic risk.

To address this gap, this survey provides a systematic review of security in LLM-assisted coding. We focus on five connected issues: the security weaknesses introduced by LLM-based generation and repair, the limitations of current evaluation datasets and pipelines, the changing threat model created by agentic and tool-integrated coding systems, the dual role of human oversight as both safeguard and risk factor, and the governance conditions required for trustworthy deployment. Thus, this review makes three main contributions. First, it synthesises security issues across code generation and repair, human interaction, and adversarial attacks, extending the analysis to the workflow-level risks of agentic and tool-integrated systems. Second, it critically examines evaluation datasets, methodologies, and metrics, with particular attention to benchmark contamination, the limitations of individual analysis methods, and the gap between benchmark settings and deployment conditions. Third, it analyses the role and limitations of human oversight and develops research directions connecting technical safeguards, evaluation practices, and deployment governance.

These dimensions are organised through the four-layer framework shown in Fig.~\ref{fig:layered-framework}, which links system capabilities, failure modes, evaluation pipelines, and controls. The framework provides a common structure for examining how different coding capabilities expose particular risks and how those risks can be assessed and mitigated.

The remainder of the paper is organised as follows. Section~\ref{related_works} reviews the existing literature and positions this work relative to surveys on software engineering, code security, AI safety, and agentic systems. Section~\ref{review_methodology} outlines the review methodology. Section~\ref{sec:security_issue} analyses the security issues associated with LLM-assisted coding, including the risks introduced by more autonomous agentic workflows. Section~\ref{sec:evaluating} examines current evaluation methods and benchmarks, and Section~\ref{sec:human_role} discusses human oversight and human-factor risks in LLM-assisted coding. Finally, Section~\ref{sec:future} offers practical insights for academia and industry and lays out directions for future work, before Section~\ref{sec:conclusion} concludes the paper.

\begin{table*}[h]
\centering
\caption{Comparison of closely related surveys and the scope of the present review. \checkmark~indicates substantial coverage; \(\sim\) indicates partial coverage; \texttimes~indicates that the topic is not a central focus.}
\label{tab:related-survey-comparison}
\small
\setlength{\tabcolsep}{4pt}
\renewcommand{\arraystretch}{1.15}
\scalebox{0.9}{
\begin{tabular}{p{2cm}p{0.6cm}p{2.2cm}p{0.85cm}p{0.8cm}p{0.8cm}p{0.8cm}p{0.8cm}p{3.8cm}}
\toprule
\textbf{Survey} & \textbf{Year} & \textbf{Primary scope} & \textbf{Sec} & \textbf{Eval.} & \textbf{Human Factor} & \textbf{Agentic} & \textbf{Gov.} & \textbf{Main gap} \\
\midrule

Huang et al.~\cite{huang2023survey} & 2023 & APR broadly & \texttimes & \texttimes & \texttimes & \texttimes & \texttimes & Not LLM-security focused. \\
Anand et al.~\cite{anand2024comprehensive} & 2024 & APR+code gen. & \(\sim\) & \(\sim\) & \texttimes & \texttimes & \texttimes & Broad overview; limited security synthesis. \\
Zhou et al.~\cite{zhou2024large} & 2024 & Vulnerability detect./repair & \checkmark & \(\sim\) & \texttimes & \texttimes & \texttimes & Task-specific rather than integrated. \\
Zhang et al.~\cite{zhang2024systematic} & 2024 & LLMs for APR & \(\sim\) & \(\sim\) & \texttimes & \texttimes & \texttimes & Repair-focused; narrow security scope. \\
Basic and Giaretta~\cite{basic2025vulnerabilities} & 2025 & LLMs in code security & \checkmark & \(\sim\) & \texttimes & \texttimes & \texttimes & Limited on deployment and governance. \\
Sheng et al.~\cite{sheng2025llms} & 2025 & Vulnerability detection & \checkmark & \checkmark & \texttimes & \texttimes & \texttimes & Detection-oriented only. \\

Li and Chow~\cite{chow2025large} & 2025 & AI safety/governance & \(\sim\) & \texttimes & \checkmark & \(\sim\) & \checkmark & Not coding-task specific. \\
Li and Chow~\cite{chow2026dialogue} & 2026 & Agentic AI systems & \(\sim\) & \texttimes & \checkmark & \checkmark & \checkmark & Not code-security centered. \\
\textbf{This survey} & \textbf{2026} & \textbf{Security in LLM-assisted coding} & \checkmark & \checkmark & \checkmark & \checkmark & \checkmark & \textbf{Integrated security-centered view.} \\
\bottomrule
\end{tabular}
}
\end{table*}

\section{Related Works}
\label{related_works}
Existing surveys explore the integration of LLMs into software engineering tasks like Automated Program Repair (APR) and code generation, increasingly focusing on security implications. Anand et al.~\cite{anand2024comprehensive} review 27 recent papers on LLMs in APR and code generation, noting their transformative impact in reducing manual debugging and improving code generation. It contrasts methodologies, identifies trends like using LLMs and feedback loops, and discusses challenges, including functional correctness and security. Similarly, Huang et al.~\cite{huang2023survey} provide a broad overview of APR techniques and discuss the opportunities introduced by learning-based methods, including LLMs.
More recently, Zhang et al.~\cite{zhang2024systematic} present a systematic literature review focused specifically on LLM applications in APR, analysing repair strategies, utilisation patterns, and repair scenarios that include security vulnerabilities.

Other surveys narrow their focus specifically to the intersection of LLMs and code security. Basic and Giaretta~\cite{basic2025vulnerabilities} investigate security vulnerabilities introduced by LLM-generated code, evaluate LLM effectiveness in detecting and fixing vulnerabilities, and explore data poisoning impacts. It aims to provide an in-depth analysis across these three areas. Likewise, Sheng et al.~\cite{sheng2025llms} survey LLM-based software vulnerability detection, examining problem formulations, models, datasets, and evaluation metrics, while Zhou et al.~\cite{zhou2024large} review the use of LLMs for vulnerability detection and repair and outline future research directions. These studies provide useful foundations, but they often concentrate on specific security tasks, such as detection or repair, without fully integrating broader issues such as benchmark reliability, human-AI trust, deployment governance, or the security implications of increasing model autonomy.

A growing body of broader LLM safety research examines issues that intersect closely with secure code generation, including controllability, misuse resistance, human oversight, and deployment governance. Li and Chow~\cite{chow2025large} help frame risks that extend beyond insecure code artifacts, such as trust miscalibration, weak oversight, and insufficient accountability. These perspectives complement code-centric security analysis by explaining how flawed outputs become operational and organizational risks. Another emerging direction concerns agentic AI systems. Coding systems are increasingly evolving from passive assistants into tool-using agents that can plan, retrieve context, execute tests, and revise files. In another work, Li and Chow~\cite{chow2026dialogue} show that this shift expands the risk landscape from single-output vulnerabilities to workflow-level risks such as tool misuse, persistent memory, compromised retrieval, and compounding multi-step errors.

Table~\ref{tab:related-survey-comparison} summarizes the main scope and limitations of closely related surveys. Prior reviews tend to focus on APR, vulnerability detection and repair, or broader AI safety questions separately. By contrast, this survey provides an integrated security-centered perspective on LLM-assisted coding across generation, repair, evaluation, human oversight, agentic risk, and deployment governance. We also place particular emphasis on evaluation limitations, including benchmark contamination, the partial visibility of static or dynamic analysis in isolation, and the mismatch between function-level benchmarks and real-world deployment settings.

Accordingly, this review is intended not only to summarize vulnerabilities and defences, but also to clarify how technical, human, and organizational factors jointly shape the secure use of LLM-assisted coding systems.

\section{Review Methodology}
\label{review_methodology}
To provide a comprehensive overview of the security landscape of LLM-based code tasks, a systematic literature review (SLR) was conducted. The methodology was designed following established guidelines for systematic reviews in software engineering and consists of three main stages: defining the research questions, executing a structured search and selection strategy, and performing a final snowballing search to ensure completeness.

\begin{figure}[t]
    \centering
    \includegraphics[width=1\linewidth]{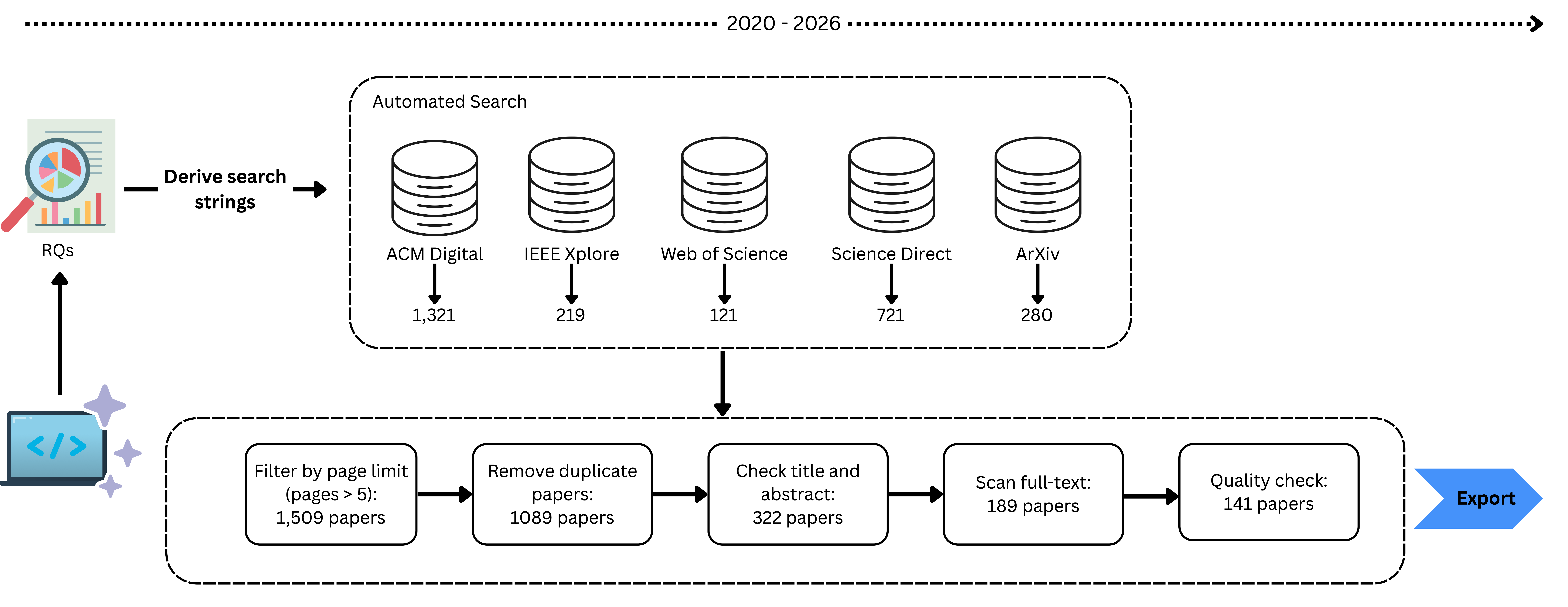}
    \vspace{-1cm}
    \caption{Search process} 
\label{fig:search_process} 
\vspace{-0.5cm}
\end{figure}

\subsection{Research Questions}
This review is guided by the following research questions (RQs), which address the core objectives identified in the introduction:
\begin{itemize}
    \item \textbf{RQ1:} What are the primary security issues introduced by the use of LLMs in code-related tasks?
    \item \textbf{RQ2:} What are the capabilities and limitations of current evaluation methods and benchmarks for assessing LLM-assisted coding tasks?
    \item \textbf{RQ3:} What is the role of human oversight and interaction in mitigating security risks in LLM-assisted software development?
\end{itemize}

\subsection{Search Strategy}
A systematic search strategy was designed to identify relevant literature, as shown in Fig.~\ref{fig:search_process}. The search was conducted across several major digital libraries and preprint archives, including ACM Digital Library, IEEE Xplore, SpringerLink, ScienceDirect, Web of Science, and arXiv, reflecting common sources used in similar comprehensive reviews.
The search keywords were formulated to capture the intersection of LLMs, code-related tasks, and security. Boolean operators were used to combine terms from three main categories:
\begin{itemize}
    \item LLM terms: "Large Language Model", "LLM", "Foundation Model", "Generative AI".
    \item Code task terms: "code generation", "program synthesis", "code completion", "program repair", "vulnerability detection", "vulnerability fixing", "automated program repair", "APR", "code analysis", "software development".
    \item Security terms: "security", "vulnerability", "secure code", "insecure code", "exploit", "attack", "threat", "prompt injection", "jailbreak", "data poisoning", "adversarial attack", "data leakage", "confidentiality", "integrity", "robustness". 
    \item Human-AI interaction terms: "human-in-the-loop", "human oversight", "human-AI collaboration", "cognitive bias", "feedback", "human review"
\end{itemize}

The timeframe for the search initially focused on publications from 2020 to early 2025, aligning with the emergence and rapid proliferation of LLMs in software engineering tasks, particularly the significant surge in research from 2023 onwards. To ensure comprehensive coverage, both forward and backward snowballing techniques were applied to the reference lists of selected key papers, a method also employed in other systematic reviews~\cite{zhang2024systematic}.

\subsection{Study Selection Criteria}
The initial search results were subjected to a rigorous, multi-stage filtering process to determine inclusion in the final review.

First, titles, abstracts, and keywords were screened for relevance. Subsequently, the full texts of the remaining articles were reviewed against the predefined inclusion and exclusion criteria.

Inclusion Criteria:
\begin{itemize}
    \item The paper must discuss security aspects directly related to the use of LLMs in tasks involving source code (generation, analysis, modification, repair, etc.).
    \item The paper evaluates LLMs for security-related tasks performed on source code (e.g., vulnerability detection, patch generation).
    \item The paper proposes or analyses adversarial attacks against LLMs specifically within the context of code-related tasks, or defences against such attacks.
    \item The study must be a peer-reviewed conference paper or journal article. Exceptionally, recent high-quality preprints were retained only when they addressed fast-moving topics not yet represented by an archival version and when their methodology and evidence were sufficiently detailed for assessment. 
\end{itemize}

Exclusion Criteria:
\begin{itemize}
    \item Papers where LLMs are not a central component of the discussed code task or where the security aspect is trivial or out of scope.
    \item Papers not written in English.
    \item Editorials, opinion pieces, abstracts, posters, or very short workshop papers lacking substantial methodological detail or findings.
    \item Studies focusing on the security of LLMs in purely natural language processing tasks (e.g., general text generation, chatbots for non-technical domains) unless the principles or findings are directly and significantly applicable to code security.
    \item Studies focusing solely on ethical concerns (e.g., job displacement, copyright) without a direct technical security component.
\end{itemize}

\subsection{Quality Assessment of Preprints}
Because this field is evolving rapidly, a limited number of recent preprints were retained when no peer-reviewed counterpart was available and when the work addressed important emerging topics. To reduce the risk of relying on unstable evidence, we applied additional screening to these studies, considering author credibility, methodological transparency, evidence quality, and the availability of later archival versions where applicable.

\begin{table}[t!]
\caption{A Taxonomy of LLM Security Issues}
\label{tab:llm_security}
\scalebox{0.85}{
\begin{tabular}{p{2.2cm} p{3.2cm} p{5.4cm} p{4.2cm}}
    \toprule
    \textbf{Category} & \textbf{Specific Threat} & \textbf{Potential Impact} & \textbf{References} \\
    \midrule

    \multirow{4}{=}{\parbox{2.1cm}{Inherent Security Challenges}} 
    & Trained with risk datasets & Introduction of exploitable flaws into applications. & ~\cite{nazzal2024promsec, mohsin2024can, gong2024well, khoury2023secure, pearce2022asleep, goetz2024you, tihanyi2025secure} \\ \cmidrule(l){2-4} 
    \addlinespace
    
    & Lack of security principle understanding & Subtle, hard-to-detect vulnerabilities; insecure design patterns. & ~\cite{mohsin2024can, asare2023github, gong2024well, he2024instruction, hajipour2024codelmsec,khoury2023secure, toth2024llms} \\ \cmidrule(l){2-4} 
    \addlinespace
    & Insecure/incomplete patches & Regression of security posture; false sense of security. & ~\cite{belozerov2025secure, tihanyi2025secure, gong2024well, asare2023github, liu2024no} \\ \cmidrule(l){2-4} 
    \addlinespace
    & Propagation of outdated/vulnerable dependencies & System compromise through known exploits in third-party components. & ~\cite{kharma2025security, pearce2022asleep, sandoval2023lost, siddiq2024sallm, mohsin2024can} \\ 
    
    \cmidrule(l){2-4} 
    \addlinespace
    & Hallucinated APIs / semantic misassumptions & Incorrect security assumptions; broken validation logic; misuse of nonexistent or misunderstood protections. & ~\cite{liu2024exploring, mohsin2024can, tie2024llms} \\
    
    \midrule

    \multirow{4}{=}{\parbox{2.1cm}{Human-Related Security Issues}} 
    & Insecure prompt engineering & Generation of vulnerable code; exposure of sensitive information. & ~\cite{goetz2024you, nazzal2024promsec, elgedawy2024ocassionally, chen2025assessing, li2024exploratory, he2024instruction} \\ \cmidrule(l){2-4} 
    \addlinespace
    & Over-reliance and reduced scrutiny & Insecure code integrated into production systems. & ~\cite{perry2023users, belozerov2025secure, hamer2024just, tihanyi2024neutral} \\ \cmidrule(l){2-4} 
    \addlinespace
    & Lack of developer security training for LLM usage & Increased likelihood of insecure code generation and deployment. & ~\cite{mohsin2024can, khoury2023secure, kirova2024software} \\
    \midrule

    \multirow{4}{=}{\parbox{2.1cm}{Attack-Related Security Issues}} 
    & Adversarial Attacks/Data Poisoning & Incorrect security assessments; introduction of exploitable flaws or backdoors. & ~\cite{xie2020adversarial, duan2021advdrop, basic2024large, zeng2025inducing, he2023large, schuster2021you, cotroneo2024vulnerabilities, yan2024llm, oh2024poisoned,aghakhani2024trojanpuzzle} \\ \cmidrule(l){2-4} 
    \addlinespace
    & Data Leakage (from training data or prompts) & Exposure of proprietary algorithms, API keys, personal data. & ~\cite{nazzal2024promsec, tihanyi2025secure, gong2024well, rabbi2024ai, li2024exploratory, hajipour2024hexacoder, fu2024security, liu2024no, he2023large, zhong2024can, hamer2024just} \\
    \bottomrule
\end{tabular}
}
\vspace{-0.5cm}
\end{table}

\section{LLM-Assisted Coding Task Security Issues}
\label{sec:security_issue}
Integrating LLMs into software development introduces diverse security issues, including inherent technological flaws, hallucination-related semantic failures, human-driven vulnerabilities, and direct attacks on the models (summarised in Table~\ref{tab:llm_security}). This section explores these categories to address \textbf{RQ1}.

Understanding this threat landscape is necessary because weaknesses originate from multiple sources, ranging from the underlying technology to specific deployment and usage configurations. Consequently, mitigation requires a layered approach. For example, a model replicating insecure training data, a malicious data poisoning attack, and a poorly written user prompt are distinct problems that require separate defensive strategies.

\subsection{Inherent Security Challenges}
A primary concern is that LLMs frequently generate vulnerable code. Because they replicate patterns from large training datasets, they often reproduce insecure practices, common weaknesses (CWEs), and outdated library usages. Lacking a semantic understanding of security principles~\cite{das2025security}, LLMs produce code that may be functionally correct but structurally insecure, simply mirroring flaws present in their training data~\cite{basic2024large}.

Additionally, the confident presentation of LLM outputs can cause developers to accept generated code with insufficient review~\cite{mohsin2024can}, overlooking subtle, exploitable bugs~\cite{yan2024llm}. This risk also applies to automated program repair (APR) and vulnerability patching. While an LLM might resolve a targeted functional error, the resulting patch can introduce new security flaws or provide an incomplete fix that attackers can still bypass~\cite{nong2024automated,alrashedy2023can}.

\noindent\textbf{Training Data Contamination.} Several studies highlight that LLMs have an inherent tendency to introduce security vulnerabilities, often due to being trained on massive open-source code corpora, which may contain insecure practices~\cite{nazzal2024promsec, mohsin2024can}. Some research suggests this is partly due to LLMs' lack of awareness regarding scenario-relevant security risks, as they primarily prioritise functional requirements~\cite{gong2024well, khoury2023secure, pearce2022asleep}.  Consequently, LLMs replicate these security flaws from their training data, leading to the generation of vulnerable code.  This creates significant concerns about the reliability of LLMs for secure code generation.

The empirical evidence for this is compelling. Götz and Schaad~\cite{goetz2024you} found that up to 65\% of code generated by some LLMs was initially insecure, attributing this partly to a lack of best-practice examples in training data. Similarly, Pearce et al.~\cite{pearce2022asleep} notably found that approximately 40\% of code snippets generated by GitHub Copilot were susceptible to vulnerabilities. This was linked to Copilot being trained on public GitHub code known to contain insecure patterns.
Mohsin et al.~\cite{mohsin2024can} demonstrate that LLMs trained on public data consistently reproduce software vulnerabilities. Their zero-shot evaluations expose a widespread lack of baseline security knowledge across multiple models. Notably, systems like ChatGPT and Copilot produced insecure code in roughly 40\% of test cases, and baseline Prompt-Driven Code Generators (PDCGs) similarly failed to meet acceptable security standards.

More comprehensive evaluations by Tihanyi et al. Tihanyi et al.~\cite{tihanyi2025secure} across nine state-of-the-art models, showing that at least 62\% to 63\% of generated C programs were vulnerable when using neutral prompts. They concluded that all examined models notoriously introduce vulnerabilities at unacceptable rates, attributing this to training on human-generated code containing vulnerabilities.

\noindent\textbf{Limited Security Principle Understanding.} Another critical inherent challenge is rooted in the design and training objectives of LLMs, which almost universally prioritise functional correctness and linguistic fluency over security awareness~\cite{mohsin2024can}. These models are engineered to predict the most probable next token based on statistical patterns, not to reason abstractly about security implications or to evaluate potential attack vectors~\cite{asare2023github}. This architectural limitation means that the security quality of generated code is often inconsistent and unpredictable.

Gong et al.~\cite{gong2024well} observed that models often produce functional code while overlooking hidden vulnerabilities, falling into scenario-specific "traps." The authors attribute this failure to a general lack of security awareness and an inability to identify risks not explicitly detailed in the prompt.
Studies show that even security-aware instructions in prompts may not significantly enhance security~\cite{he2024instruction} or reliably~\cite{hajipour2024codelmsec,khoury2023secure} reduce vulnerabilities. T{\'o}th et al.~\cite{toth2024llms} highlighted that security risks can be highly situational, making it difficult for AI to grasp context and distinguish safe from risky practices.

\noindent\textbf{Insecure or Incomplete Patches.} The application of LLMs to automated program repair (APR) and vulnerability patching introduces another vector of risk. Fixes generated by LLMs may be functionally correct on the surface but can introduce new security flaws or leave the original vulnerability only partially mitigated. 

For example, Belozerov et al.~\cite{belozerov2025secure} demonstrate that ChatGPT and GPT-4o correctly identified and fixed only 17 of 32 tested vulnerabilities. The models often produced incomplete patches, relied on superficial mitigations such as simple input validation, and misclassified vulnerability types (e.g., confusing CWE-78~\cite{CWE-78} with CWE-77~\cite{CWE-77}), thereby applying inappropriate defenses that left the original flaws unresolved.

Tihanyi et al.~\cite{tihanyi2025secure} suggest that current LLMs are not yet ready for program repair because they still struggle to generate functionally correct fixes. Although some studies report promising results, particularly when combined with other tools, substantial further research is still needed~\cite{tihanyi2025secure, gong2024well}.

Asare et al.~\cite{asare2023github} show that Copilot may reintroduce vulnerabilities rather than produce secure repairs, while scanner-guided iterative repair eventually plateaus. Likewise, Liu et al.~\cite{liu2024no} report that repair success varies by vulnerability type, with incomplete fixes persisting for classes such as Input Validation and Information Leakage.

\noindent\textbf{Outdated Dependency Propagation.} LLMs often recommend deprecated libraries, obsolete APIs, and cryptographically broken implementations that persist in training data despite safer alternatives. Studies report risks from outdated libraries~\cite{kharma2025security} and deprecated APIs~\cite{pearce2022asleep, sandoval2023lost}, indicating that obsolete practices can continue to appear in generated code as the security landscape evolves~\cite{pearce2022asleep}. One example is MD5, a broken hash algorithm still suggested by models such as GitHub Copilot~\cite{sandoval2023lost} and also identified in code analyzed by the Sallm framework under CWE-328~\cite{siddiq2024sallm}. Because LLMs are trained on public repositories, they may reproduce outdated or vulnerable third-party dependencies in both prompt-driven models and coding assistants~\cite{mohsin2024can, yao2024survey}. While security quality varies across models, languages, and domains, severe vulnerabilities still recur under simple prompts.

\noindent\textbf{Hallucinated APIs and Security-Relevant Semantic Misassumptions.} Another inherent risk in LLM-assisted coding is hallucination: models may generate nonexistent APIs, invalid signatures, fabricated library behavior, or incorrect assumptions about security mechanisms. Although often framed as a functional correctness issue, such hallucinations can also have direct security consequences. For example, a model may invent a sanitization routine, misuse an authentication or cryptographic API, assume that a library enforces checks it does not provide, or produce plausible code that silently violates security requirements~\cite{mohsin2024can, tie2024llms}.

These errors are particularly dangerous because they can appear superficially convincing. Developers may fix syntax or compilation issues while missing deeper semantic errors involving permissions, validation, state handling, or security boundaries. Recent work shows that current models still struggle to detect hallucinated code and to repair it reliably once introduced~\cite{liu2024exploring}. In more autonomous or agentic settings, such errors can also propagate across steps when hallucinated dependencies, fabricated tool capabilities, or mistaken execution assumptions are reused as context. Hallucination should therefore be treated not just as a coding nuisance, but as a security-relevant failure mode linking model uncertainty, plausible output, and exploitability.

\begin{figure}[t]
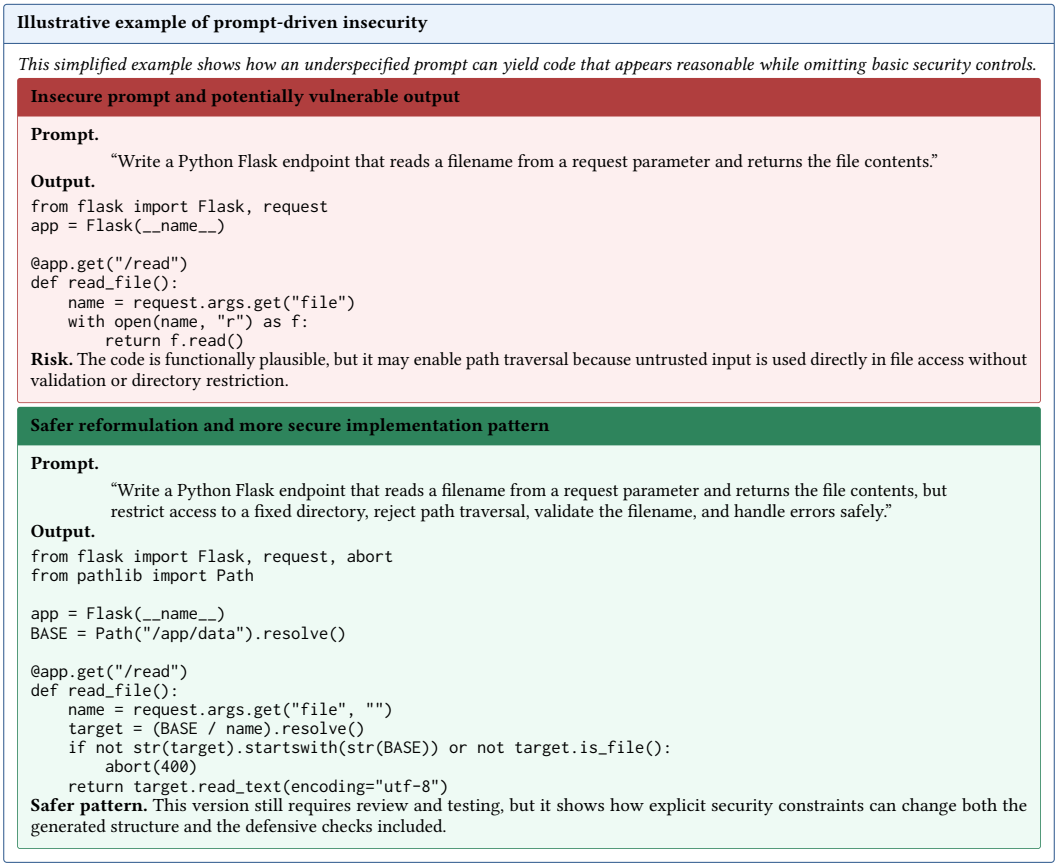

\centering
\scriptsize
\setlength{\parskip}{0pt}

\begin{tcolorbox}[
    enhanced,
    width=\linewidth,
    boxrule=0.4pt,
    arc=1pt,
    colback=white,
    colframe=titleblue,
    colbacktitle=titlebg,
    coltitle=black,
    title={\textbf{Illustrative example of prompt-driven insecurity}},
    fonttitle=\scriptsize,
    left=2pt,right=2pt,top=2pt,bottom=2pt,
    toptitle=1pt,bottomtitle=1pt,
    before skip=0pt,after skip=0pt
]

\textit{This simplified example shows how an underspecified prompt can yield code that appears reasonable while omitting basic security controls.}

\vspace{0.2em}

\begin{tcolorbox}[
    enhanced,
    boxrule=0.3pt,
    arc=0.8pt,
    colback=riskbg,
    colframe=riskred,
    title={\textbf{Insecure prompt and potentially vulnerable output}},
    coltitle=black,
    fonttitle=\scriptsize,
    left=2pt,right=2pt,top=1.5pt,bottom=1.5pt,
    toptitle=0.8pt,bottomtitle=0.8pt,
    before skip=0pt,after skip=0pt
]

\textbf{Prompt.}
\begin{quote}
\setlength{\leftskip}{0.8em}
\setlength{\rightskip}{0.8em}
\setlength{\topsep}{0pt}
\setlength{\partopsep}{0pt}
``Write a Python Flask endpoint that reads a filename from a request parameter and returns the file contents.''
\end{quote}

\vspace{-0.4em}
\textbf{Output.}

\begin{Verbatim}[fontsize=\scriptsize,baselinestretch=0.9, xleftmargin=0pt]
from flask import Flask, request
app = Flask(__name__)

@app.get("/read")
def read_file():
    name = request.args.get("file")
    with open(name, "r") as f:
        return f.read()
\end{Verbatim}

\vspace{-0.4em}
\textbf{Risk.} The code is functionally plausible, but it may enable path traversal because untrusted input is used directly in file access without validation or directory restriction.

\end{tcolorbox}

\vspace{0.2em}

\begin{tcolorbox}[
    enhanced,
    boxrule=0.3pt,
    arc=0.8pt,
    colback=safebg,
    colframe=safegreen,
    title={\textbf{Safer reformulation and more secure implementation pattern}},
    coltitle=black,
    fonttitle=\scriptsize,
    left=2pt,right=2pt,top=1.5pt,bottom=1.5pt,
    toptitle=0.8pt,bottomtitle=0.8pt,
    before skip=0pt,after skip=0pt
]

\textbf{Prompt.}
\begin{quote}
\setlength{\leftskip}{0.8em}
\setlength{\rightskip}{0.8em}
\setlength{\topsep}{0pt}
\setlength{\partopsep}{0pt}
``Write a Python Flask endpoint that reads a filename from a request parameter and returns the file contents, but restrict access to a fixed directory, reject path traversal, validate the filename, and handle errors safely.''
\end{quote}

\vspace{-0.4em}
\textbf{Output.}

\begin{Verbatim}[fontsize=\scriptsize,baselinestretch=0.9, xleftmargin=0pt]
from flask import Flask, request, abort
from pathlib import Path

app = Flask(__name__)
BASE = Path("/app/data").resolve()

@app.get("/read")
def read_file():
    name = request.args.get("file", "")
    target = (BASE / name).resolve()
    if not str(target).startswith(str(BASE)) or not target.is_file():
        abort(400)
    return target.read_text(encoding="utf-8")
\end{Verbatim}

\vspace{-0.4em}
\textbf{Safer pattern.} This version still requires review and testing, but it shows how explicit security constraints can change both the generated structure and the defensive checks included.

\end{tcolorbox}

\end{tcolorbox}

\caption{Illustrative example showing how weak prompt specification can lead to insecure but plausible code, and how a more security-aware reformulation can elicit safer implementation patterns.
The prompts were manually written, and the sample code was generated using Gemini 3.1 Pro and subsequently reviewed by the authors for illustrative accuracy.}
\label{fig:prompt-insecurity-example}
\vspace{-0.5cm}
\end{figure}

\subsection{Human-Related Security Issues}
Beyond the inherent limitations of the models themselves, the human element represents a significant source of security vulnerabilities in LLM-assisted coding. The interaction between the developer and the model, from the initial prompt to the final code review, is a critical stage where misunderstandings, biases, and a lack of security awareness can lead to the introduction of serious flaws.

\noindent\textbf{Insecure Prompt Engineering.} 
User prompts can inadvertently steer LLMs toward insecure code or bypassed safeguards. The security of generated code depends heavily on prompt quality and specificity: vague prompts often yield insecure outputs~\cite{goetz2024you}, and models may fail to infer or prioritize risks that are not explicitly stated~\cite{gong2024well}. Consequently, non-expert users may obtain code that is functionally correct yet vulnerable~\cite{nazzal2024promsec}, and even clearly security-sensitive tasks may not trigger warnings or protective measures unless these are explicitly requested~\cite{nazzal2024promsec, belozerov2025secure}.

Crafting effective prompts is therefore a non-trivial skill, requiring both clear communication and engineering judgment~\cite{elgedawy2024ocassionally}. Novice users often struggle to specify security requirements precisely enough to elicit secure code~\cite{chen2025assessing}. However, simply asking for ``secure'' code is usually insufficient; zero-shot security directives may still fail to prevent vulnerabilities such as path traversal~\cite{goetz2024you}, and generic security boilerplate alone yields little improvement~\cite{he2024instruction}. More detailed security-aware prompts can help, but only when users can accurately identify and describe the relevant requirements~\cite{li2024exploratory}. Figure~\ref{fig:prompt-insecurity-example} illustrates how underspecified prompts can produce plausible code that omits essential security controls.

\noindent\textbf{Over-Reliance and Reduced Scrutiny.} Developers excessively trust LLM outputs, leading to insufficient review of generated code.

Perry et al.~\cite{perry2023users} indicate that programmers using AI assistants produced less secure code than those working independently in four of five security-focused tasks. Yet these users were more likely to believe their code was secure, indicating overconfidence. The resulting code often omitted secure library use, proper input sanitisation, and edge-case handling for databases and file systems. 

Belozerov et al.~\cite{belozerov2025secure} caution against over-reliance on AI code generation, arguing that manual review and security testing remain essential. Because LLMs can create a false sense of safety, all generated output requires rigorous inspection. Hamer et al.~\cite{hamer2024just} also warn against blindly trusting external code, whether AI-generated or human-written.

Tihanyi et al.~\cite{tihanyi2024neutral} argue that relying on LLMs for production code generation without expert oversight is unsafe. Their evaluation of eight major models found vulnerabilities in at least 63.47\% of generated C programs, covering 42 CWEs, including six from MITRE's Top 25; dereference failures and buffer overflows were the most frequent. Even the best models produced vulnerable code in roughly half of the cases, underscoring the need for rigorous manual validation.

\noindent\textbf{Insufficient Security Training for LLM Usage.} 
Developers with limited security expertise are particularly vulnerable to misusing LLMs. Mohsin et al.~\cite{mohsin2024can} show that developer experience strongly affects the security of LLM-assisted coding: inexperienced programmers, especially in complex languages such as C++, often overlook critical practices, leading to vulnerabilities such as buffer overflows and improper input validation. Limited understanding of prompt security and subtle flaws in LLM output further increase the risk of exploitable code entering production.

Khoury et al.~\cite{khoury2023secure} reinforce this concern by simulating novice programmers, finding that ChatGPT produced initially secure code in only 5 of 21 cases without explicit security instructions. The model frequently generated severe vulnerabilities, including SQL injection, path traversal, and insecure handling of sensitive operations such as unhashed passwords, and typically added basic safeguards only when asked explicitly.

Accordingly, software engineering education must adapt as LLMs automate routine coding~\cite{kirova2024software}. Developers need stronger skills in prompt engineering, critical evaluation, and artifact verification, while curricula should emphasize review, testing, validation, and verification, together with professional ethics, responsible AI, and secure development practices~\cite{kirova2024software, mohsin2024can}. Systematic security testing and review therefore remain essential when using LLM-generated code.

\subsection{Attack-Related Security Issues}
Beyond inherent weaknesses and user-related errors, LLMs are themselves direct targets for attacks designed to compromise the integrity, confidentiality, and availability of the code they generate. These attacks exploit vulnerabilities in the LLM's architecture, its training data, or its input-processing mechanisms to maliciously influence its behaviour.

\noindent\textbf{Adversarial Attacks and Data Poisoning.} 
A primary attack vector is adversarial input, where carefully crafted prompts bypass an LLM's safety mechanisms~\cite{shayegani2023survey}. Successful prompt injection can induce harmful code generation, confidential data disclosure, or unintended actions in integrated environments. As in adversarial attacks on image models~\cite{xie2020adversarial, duan2021advdrop}, subtle input modifications can also make coding LLMs misclassify vulnerable code as safe, suggest flawed patches, or miss existing vulnerabilities. More broadly, poisoning the data used for training or fine-tuning can implant backdoors or systemic weaknesses that later reappear in generated code~\cite{cotroneo2024vulnerabilities, tessa2026secure, basic2024large}.

Zeng et al.~\cite{zeng2025inducing} demonstrate this risk through HACKODE, which hides adversarial sequences in external code references, often within comments. Although these sequences appear meaningless to humans, they can alter model outputs while leaving the infected examples functionally correct, allowing them to spread through highly ranked online sources and into coding assistants. As a result, models may generate unsafe code from seemingly legitimate inputs, making the attack difficult to detect.

Poisoning attacks are also a serious training-phase threat. He et al.~\cite{he2023large} show that poisoned data or malicious fine-tuning can bias code models toward insecure cryptographic recommendations and distinguish these attacks from SVEN, their post-training evaluation method. Schuster et al.~\cite{schuster2021you} further classify poisoning threats into model poisoning, which compromises the model or its supply chain, and data poisoning, which inserts malicious files into training repositories.

Some poisoning attacks inject explicit insecure payloads into training data~\cite{cotroneo2024vulnerabilities, yan2024llm, oh2024poisoned, schuster2021you}, but such payloads may be detectable by static analysis. To evade this, Aghakhani et al.~\cite{aghakhani2024trojanpuzzle} propose COVERT, which hides malicious content in ignored regions such as comments and docstrings, and TROJANPUZZLE, which teaches the model token substitution patterns without including the full payload directly. When triggered, the model can then reconstruct and emit the malicious payload automatically.

\noindent\textbf{Data Leakage and Confidentiality Breaches.}  
LLM-generated code often contains vulnerabilities that can expose, alter, or destroy sensitive data, thereby compromising confidentiality and sometimes availability. Attacker-induced flaws such as buffer overflows can also enable unauthorised access or modification of critical information~\cite{nazzal2024promsec, zeng2025inducing}.

Improper input validation (CWE-20~\cite{CWE-20}) and related injection flaws, including SQL injection (CWE-89~\cite{CWE-89}), cross-site scripting (CWE-79~\cite{CWE-79}), OS command injection (CWE-78~\cite{CWE-78}), and code injection (CWE-94~\cite{CWE-94}), are common~\cite{tihanyi2025secure, fu2024security, gong2024well, rabbi2024ai}. These vulnerabilities can enable data theft, tampering, deletion, or even system compromise~\cite{nazzal2024promsec}. A common cause is missing input validation or filtering, for example, failing to sanitize file names and thereby enabling path traversal (CWE-22~\cite{CWE-22})~\cite{liu2024no, toth2024llms}.

Memory-safety flaws, such as buffer overflows (CWE-120~\cite{CWE-120}, CWE-787~\cite{CWE-787}, CWE-119~\cite{CWE-119}), out-of-bounds reads or writes (CWE-125~\cite{CWE-125}, CWE-787~\cite{CWE-787}), null pointer dereferences (CWE-476~\cite{CWE-476}, CWE-153~\cite{CWE-153}, CWE-346~\cite{CWE-346}), and memory leaks or use-after-free errors (CWE-401~\cite{CWE-401}, CWE-416~\cite{CWE-416}) are also frequent, especially in C/C++~\cite{tihanyi2024neutral, li2024exploratory, liu2024no, hajipour2024hexacoder}. These may cause crashes, data leakage, or arbitrary code execution, and missing NULL checks after allocation are a recurring weakness~\cite{liu2024no}.

Information exposure vulnerabilities (e.g., CWE-200~\cite{CWE-200}, CWE-352~\cite{CWE-352}) are likewise reported~\cite{fu2024security, mohsin2024can}. A common pattern is hard-coded credentials, where models generate passwords or API keys as clear-text strings~\cite{gong2024well}. Invalid validation can also lead to unsafe operations that leak sensitive information~\cite{liu2024no}.

Further risks include API misuse, such as insecure cryptographic choices (CWE-327~\cite{CWE-327}, CWE-759~\cite{CWE-759}) or unsafe deserialization (CWE-502~\cite{CWE-502}), as well as resource leaks~\cite{he2023large, fu2024security, zhong2024can}. Other reported issues include path traversal, unrestricted file upload, incorrect permissions, unsafe dynamic execution, and infinite loops that may trigger denial-of-service conditions~\cite{CWE-22, CWE-434, CWE-732, CWE-94, CWE-835, zeng2025inducing, hamer2024just, gong2024well, liu2024no}.

\subsection{Security Implications of Agentic Coding Systems}

Early studies often evaluate LLM-assisted coding as a largely standalone task, but recent research and practice are shifting toward \emph{agentic coding systems}: autonomous or semi-autonomous LLM-based systems that can decompose tasks, invoke tools, retrieve repository context, execute tests, revise files, and iterate on intermediate results~\cite{chow2026dialogue, nunez2024autosafecoder, liu2024agents4plc, zheng2024opencodeinterpreter}. This shift changes the threat model from a single-output artifact problem to a multi-step socio-technical workflow problem.

First, agentic systems expand the attack surface because they interact with shells, package managers, testing frameworks, retrieval systems, external APIs, and local development environments~\cite{nunez2024autosafecoder, liu2024agents4plc}. Prompt injection may therefore arise not only from user input, but also from retrieved documentation, repository comments, malicious tool output, poisoned test artifacts, or adversarial external resources~\cite{chow2026dialogue, zheng2024opencodeinterpreter}. As these components multiply, so do the trust boundaries that must be secured.

Second, agentic systems can amplify errors across steps. Unlike a standalone insecure completion, an early hallucination, false assumption, or unsafe action may cascade by selecting a vulnerable dependency, generating flawed tests, approving an incomplete patch, or overwriting secure code with insecure revisions~\cite{chow2026dialogue}. Because later stages may reuse earlier outputs as context, the system can recursively propagate its own mistakes.

Third, agentic systems can reduce human visibility and controllability. Developers may see only the final artifact rather than the intermediate reasoning, tool calls, retrieved evidence, and edits that produced it, making it harder to audit vulnerabilities, assign responsibility, or identify whether failures stem from the base model, a tool interaction, a retrieval source, or workflow design~\cite{liu2024agents4plc}. Secure deployment therefore requires not only code inspection but also traceability across the broader execution process.

Finally, agentic coding systems blur the boundary between software security and broader AI safety concerns, including controllability, misuse resistance, and operational governance~\cite{nunez2024autosafecoder, zheng2024opencodeinterpreter}. Accordingly, evaluating future LLM-assisted coding systems requires workflow-aware, repository-level, and action-oriented assessment rather than only single-turn code generation benchmarks~\cite{chow2026dialogue, liu2024agents4plc}. Taken together, these observations help explain why recurring vulnerability families appear across models, languages, prompting settings, and evaluation designs, as summarized in Table~\ref{tab:vulnerability-synthesis}; this complements Table~\ref{tab:llm_security} by shifting attention from broad risk categories to recurrent security weaknesses.

\begin{table*}[t]
\centering
\caption{Recurring vulnerability families in LLM-assisted coding studies.}
\label{tab:vulnerability-synthesis}
\scalebox{0.87}{
\begin{tabular}{p{2.6cm}p{2.6cm}p{2.6cm}p{6.8cm}}
\toprule
\textbf{Vul. Family} & \textbf{Representative CWEs} & \textbf{Representative Works} & \textbf{Recurring Pattern} \\
\midrule
Input validation and injection & CWE-20, 22, 78, 79, 89, 94 & \cite{fu2024security, gong2024well, khoury2023secure, liu2024no, nazzal2024promsec, rabbi2024ai, tihanyi2025secure, toth2024llms} & Models often generate functionally plausible code that omits sanitization, boundary checks, or command/query validation. \\
\cmidrule(l){1-4}
\addlinespace 
Memory safety & CWE-119, 120, 125, 416, 476, 787 & \cite{hajipour2024hexacoder, li2024exploratory, liu2024no, tihanyi2024neutral, tihanyi2025secure} & Memory-unsafe languages show recurring low-level errors, especially when prompts do not explicitly encode safety constraints. \\
\cmidrule(l){1-4}
\addlinespace 
Weak crypto and insecure dependencies & CWE-327, 328, 759 & \cite{he2023large, kharma2026security, mohsin2024can, pearce2022asleep, sandoval2023lost, siddiq2024sallm} & Models reuse outdated algorithms, deprecated APIs, and vulnerable dependencies inherited from training corpora. \\
\cmidrule(l){1-4}
\addlinespace 
Incomplete or insecure repair & CWE-dependent & \cite{alrashedy2023can, asare2023github, belozerov2025secure, liu2024no, nong2024automated, pearce2023examining, tihanyi2025secure} & Generated patches may satisfy local tests while leaving root causes unresolved or introducing new flaws. \\
\cmidrule(l){1-4}
\addlinespace 
Data leakage and exposure & CWE-200 and related patterns & \cite{fu2024security, gong2024well, hajipour2024hexacoder, hamer2024just, he2023large, li2024exploratory, mohsin2024can, nazzal2024promsec, zeng2025inducing, zhong2024can} & Insecure handling of secrets, credentials, and sensitive data remains a recurrent failure mode. \\
\cmidrule(l){1-4}
\addlinespace 
Hallucination-related semantic failure & Non-CWE-specific & \cite{liu2024exploring, mohsin2024can, tie2024llms} & Fabricated APIs or false assumptions about security behavior can produce plausible yet unsafe code and may propagate across workflows. \\
\bottomrule
\end{tabular}
}
\end{table*}

\section{Evaluating the Security of LLM-assisted Coding}
\label{sec:evaluating}
Assessing the security of LLM-assisted coding remains difficult because current benchmarks and evaluation pipelines capture different aspects of risk~\cite{zhou2024large, fan2023large, lyu2024automatic, li2023cctest}. Some focus on functional correctness, others probe known vulnerability patterns, and still others evaluate runtime behavior, repository context, or human review~\cite{chen2026hardsecbench, basic2024large, sheng2025large, li2025safegenbench, chen2025secureagentbench}. In addressing \textbf{RQ2}, this section examines the main datasets, metrics, and evaluation methods used in the literature, with particular attention to benchmark contamination, limited realism, and cross-framework incomparability.

\subsection{Datasets}
A crucial component of any evaluation framework is the dataset used for assessment. In the context of LLMs for code generation, these datasets can be broadly divided into those that measure functional correctness and those specifically designed to probe for security vulnerabilities.

\subsubsection{Functional Correctness}
The initial wave of benchmarks for LLM-based coding tasks focused primarily on functional correctness, whether the generated code produces the correct output for a given set of inputs. A foundational benchmark in this area is HumanEval~\cite{chen2021evaluating}, a collection of hand-written Python programming problems where the model must synthesise a function from its docstring. This benchmark has been extended to enhance its rigour and scope; HumanEval+~\cite{liu2023your} augments the original with 80 times more test cases, while HumanEval-X~\cite{zheng2023codegeex} adapts the problems for multiple languages, including C++, Java, and Go.

Austin et al.~\cite{austin2021program} introduced two complementary datasets for code synthesis: MBPP, with 974 entry-level Python tasks including a hand-verified subset of 426, and MathQA-Python, with 23,914 mathematical word problems adapted from MathQA. MBPP targets imperative algorithmic construction, whereas MathQA-Python emphasizes language understanding through straight-line code generation. Hendrycks et al.~\cite{hendrycks2021measuring} proposed APPS, a 10,000-task benchmark for Python code generation that spans introductory to competition difficulty and evaluates functional correctness through test execution.

Du et al.~\cite{du2023classeval} introduced ClassEval, the first benchmark for class-level code generation, with 200 Python tasks requiring complete classes generated from natural language descriptions. Unit-test results show that class-level generation is substantially harder than function-level generation.

For more realistic and domain-specific assessment, DS-1000 provides 1,000 Stack Overflow data-science tasks requiring major Python libraries~\cite{lai2023ds}, BigCodeBench offers 1,140 practical Python tasks involving multi-library function use~\cite{zhuo2024bigcodebench}, and CoderEval evaluates 460 repository-level Python and Java tasks requiring full-project context~\cite{yu2024codereval}. Together, these benchmarks show that practical and context-dependent generation remains challenging for current models.

\subsubsection{Security-Specific}
As the functional limitations of LLMs became better understood, a new class of benchmarks emerged to address the critical issue of security in the code they generate. These security-specific datasets are not designed to evaluate if the code works, but rather if it is safe and free from known vulnerabilities.

Early evaluation methods relied on curated prompt sets to test for known insecure patterns. For instance, SecurityEval~\cite{siddiq2022securityeval} provides 130 Python prompts mapped to 75 Common Weakness Enumeration (CWE) categories. These samples directly assess whether a model's generated output introduces specific security vulnerabilities.
Similarly, LLMSecEval\cite{tony2023llmseceval} offers 150 natural language prompts to assess code generation in vulnerability-prone scenarios. Covering 18 of MITRE's Top 25 CWEs~\cite{CWE-top-25}, these prompts were derived from secure code snippets and manually refined to be language-agnostic. To support robust comparative evaluations, the dataset includes a secure, CodeQL-verified~\cite{codeql} Python baseline for each prompt.

Wang et al.~\cite{wang2024your} proposed CodeSecEval for evaluating both code generation and repair. The dataset contains 180 Python samples spanning 44 critical vulnerability types, each with executable secure and insecure versions and test cases for automated functional and security assessment. It comprises SecEvalBase, which extends SecurityEval~\cite{siddiq2022securityeval}, and SecEvalPlus, which targets the most common and severe weaknesses.

FormAI~\cite{tihanyi2023formai} instead combines large-scale code generation with formal verification. It contains 112,000 C programs generated by GPT-3.5-turbo and labeled with ESBMC, which identifies both vulnerability type and location. The study reported vulnerabilities in 51.24\% of generated programs.

Tihanyi et al.~\cite{tihanyi2024neutral} later introduced FormAI-v2, a larger set of 331,000 C programs generated by models such as GPT-4 and Gemini-1.0-Pro. Evaluating code from neutral prompts, they found with ESBMC that at least 63.47\% of programs were vulnerable, suggesting that such risks are inherent to current code-generation models.

\subsection{Metrics}
To quantify the performance of LLMs in coding tasks, researchers employ a range of metrics. These can be divided into two main categories: those that assess the functional correctness of the generated code and those specifically designed to measure its security posture.

\subsubsection{Functional correctness}
\noindent\textbf{Pass@k:} A widely adopted metric for functional correctness is pass@k, which measures the probability that at least one of k code samples generated for a given task successfully passes all of its unit tests~\cite{chen2021evaluating, chen2025assessing, fu2024constrained, tian2023chatgpt, wang2024your, 10606356, siddiq2024sallm, dilgren2025secrepobench, mathews2024test, miah2024user}. This metric provides an indication of the likelihood that a developer will find a correct solution within a set of k generated options~\cite{wang2024your, siddiq2024sallm}. To ensure statistical robustness and reduce variance in results, an unbiased estimator of pass@k is commonly utilised in evaluations~\cite{tong2024codejudge, wang2024your, he2023large}. Some frameworks have also introduced variations, such as the ranked pass@k (pass@k@m) used by TiCoder~\cite{10606356}, which denotes correctness after a specified number of user queries and subsequent pruning of options.

\noindent\textbf{Compilation and Unit Tests:} At a fundamental level, functional correctness is determined by whether the generated code compiles successfully and subsequently passes a predefined suite of unit tests~\cite{chen2021evaluating, fu2024constrained, liu2023your, kharma2025security}. The percentage of code snippets that are merely compilable can also serve as a basic indicator of a model's output quality.

\noindent\textbf{Text Similarity Metrics:} Metrics that assess the token-level similarity between generated code and a reference solution—such as BLEU, ROUGE-L, CodeBLEU, METEOR, ChrF, and RUBY—have also been applied~\cite{ren2020codebleu, chen2021evaluating}. However, these metrics are often considered unreliable for evaluating code. Functionally equivalent programs can exhibit significant syntactic variations, meaning a high similarity score does not necessarily indicate functional correctness, nor does a low score confirm its absence~\cite{tong2024codejudge}.

\noindent\textbf{Task-Specific Criteria:} Generic metrics are often inadequate for specialised tasks, which require tailored evaluation criteria. For example, path-planning code is assessed by whether it finds valid paths and satisfies task constraints~\cite{chen2025assessing}, while hallucination detection uses dedicated metrics such as Valid Rate (VR), hallucination existence accuracy (Accrec), and hallucination type accuracy (Acctype)~\cite{liu2024exploring}.

\subsubsection{Security-Specific}
\noindent\textbf{Security Rate (SVEN-SR):} An early metric proposed for security is the SVEN-SR, defined as the percentage of secure programs among all unique and compilable generated programs~\cite{fu2024constrained}. While this metric provides a baseline indication of security, it has been criticised for potentially overestimating a model's practical security performance. By not accounting for functional correctness, a trivially safe but entirely useless piece of code could achieve a high SVEN-SR score.

\noindent\textbf{Secure-pass@k:} To provide a more realistic assessment, the Secure-pass@k metric measures the probability of generating code that is both secure and functionally correct within k attempts~\cite{fu2024constrained, dilgren2025secrepobench}. It is considered a more practical measure of an LLM's secure code generation capability because it integrates functional correctness as an essential prerequisite for a successful outcome.

\noindent\textbf{Secure@kpass:} A related but distinct metric is Secure@kpass, which assesses how likely any one out of $k$ functionally correct generations is also secure~\cite{fu2024constrained}. This helps to decouple the security evaluation from the model's ability to generate working code in the first place.

\noindent\textbf{func-sec@k:} Similarly, the func-sec@k metric evaluates functionality and security simultaneously, indicating the likelihood that an LLM-generated implementation (out of $k$ options) is both correct and secure~\cite{peng2025cweval}. This metric is particularly effective at highlighting the significant gap that often exists between a model's ability to generate functional code and its ability to generate code that is both functional and secure.

\noindent\textbf{vulnerable@k:} Approaching the problem from the opposite perspective, the vulnerable@k metric measures the probability that at least one code snippet out of k generated samples contains a detectable vulnerability~\cite{siddiq2024sallm}. In this case, a lower score for vulnerable@k signifies a better performance in generating secure code.

\noindent\textbf{False Refusal Rate (FRR):} To quantify the inherent safety-utility tradeoff in models that have been fine-tuned for safety, the FRR measures the percentage of benign prompts that an LLM incorrectly refuses because it mistakes them for unsafe or malicious requests~\cite{bhatt2024cyberseceval}.

\noindent\textbf{Static Security Analysis Tools:} Underlying many of these security metrics is the use of static security analysis tools. Tools such as CodeQL, SonarQube, Bandit, ESLint, and Infer are widely employed to automatically detect security flaws and known vulnerabilities directly within the source code~\cite{pearce2022asleep, bhatt2023purple, nunez2024autosafecoder, peng2025cweval, hajipour2024codelmsec, fu2024constrained, dolcetti2024helping, tony2023llmseceval, siddiq2024sallm, fu2024security, kharma2025security}. These tools are instrumental in identifying specific vulnerability classes and flagging security hotspots. It is important to note, however, that static analysers are known to suffer from imprecision, including both false positives and false negatives~\cite{wang2024your, li2025everything}.

\subsection{Evaluation Tools}
Security evaluations use complementary oracles. Static analyzers scale to many samples and vulnerability classes but have false positives, false negatives, and limited access to runtime or repository context~\cite{hajipour2024codelmsec, li2025everything, siddiq2024sallm, wang2024your}. Dynamic tests, fuzzing, and exploit execution provide behavioral evidence, but their conclusions depend on test coverage, reproducible environments, and the fidelity of the attack scenario~\cite{dilgren2025secrepobench, nunez2024autosafecoder, peng2025cweval, vero2025baxbench, yang2024seccodeplt}. Repository-aware evaluation exposes integration failures, while expert or developer review captures aspects of usability and oversight that automated tests cannot measure~\cite{10606356, li2025everything, miah2024user, yu2024codereval, zheng2025humanevo}. For this reason, results from different tools should be interpreted in terms of their task, oracle, context, and security objective; a fuller catalogue of the surveyed frameworks appears in Supplementary Section 2. 

\begin{table*}[t]
\centering
\caption{Representative families of evaluation frameworks for LLM-assisted coding security.}
\label{tab:framework-comparison}
\scalebox{0.86}{
\begin{tabular}{p{3cm}p{2cm}p{2.6cm}p{7cm}}
\toprule
\textbf{Framework Family} & \textbf{Representative Works} & \textbf{Primary Oracle} & \textbf{Key Use\&limitation} \\
\midrule
Function-level correctness & \cite{chen2021evaluating, austin2021program, hendrycks2021measuring} & Unit tests & Useful for baseline synthesis ability, but weak in security realism and contextual coverage. \\
\cmidrule(l){1-4}
\addlinespace 
Security prompt benchmarks & \cite{siddiq2024sallm, siddiq2022securityeval, tony2023llmseceval, wang2024your} & Curated prompts + static rules & Effective for probing known vulnerability classes, but limited in runtime realism and sensitive to analyzer imprecision. \\
\cmidrule(l){1-4}
\addlinespace 
Dynamic/exploit-driven security & \cite{dilgren2025secrepobench, nunez2024autosafecoder, peng2025cweval, vero2025baxbench, yang2024seccodeplt} & Runtime tests, fuzzing, exploit execution & Better reflects exploitability in practice, but depends heavily on test quality, environment fidelity, and setup cost. \\
\cmidrule(l){1-4}
\addlinespace 
Repository/context-aware evaluation & \cite{li2025everything, yu2024codereval, zheng2025humanevo} & Project tests + contextual reasoning & Captures integration and cross-file failures, but is harder to standardize and compare across models. \\
\cmidrule(l){1-4}
\addlinespace 
Human-centered/manual evaluation & \cite{10606356, miah2024user, perry2023users} & Expert review/user study & Reveals review burden, usability, and trust effects, but is costly, subjective, and hard to scale. \\
\bottomrule
\end{tabular}
}
\vspace{-0.5cm}
\end{table*}

\subsection{Cross-Framework Comparison and Evaluation Gaps}

The surveyed evaluation frameworks differ not only in implementation but also in what they treat as evidence of security. Some benchmarks focus on function-level correctness under predefined tests, while others probe for vulnerability patterns, repository integration, exploitability, or the interaction between human reviewers and generated code. As a result, two studies can evaluate the same model yet reach different conclusions because they observe different failure surfaces rather than because one of them is necessarily incorrect.

A useful way to compare the literature is along six dimensions: \emph{(i)} task granularity, ranging from isolated functions to repository-level workflows; \emph{(ii)} oracle type, including unit tests, static analyzers, exploit execution, and manual review; \emph{(iii)} evaluation objective, such as correctness, vulnerability detection, secure repair, or attack helpfulness; \emph{(iv)} context availability, from prompt-only generation to cross-file and temporal repository context; \emph{(v)} security realism, ranging from rule-based checks to exploit-driven validation; and \emph{(vi)} human involvement, from fully automated pipelines to expert review. These dimensions help explain why benchmark results are often not directly comparable. Table~\ref{tab:framework-comparison} summarizes these framework families and highlights the distinct trade-offs that make cross-study results difficult to compare directly.

Static-analysis-centered frameworks such as SALLM~\cite{siddiq2024sallm}, CodeLMSec~\cite{hajipour2024codelmsec}, and parts of CodeSecEval~\cite{wang2024your} scale well and provide broad CWE coverage, but they inherit the incompleteness and context limitations of the underlying analyzers. Dynamic and exploit-driven approaches such as SecRepoBench~\cite{dilgren2025secrepobench}, CWEVAL~\cite{peng2025cweval}, AutoSafeCoder~\cite{nunez2024autosafecoder}, and BAXBENCH~\cite{vero2025baxbench} better capture runtime failures and real exploitability, but they depend heavily on test quality, environment setup, and scenario design. Repository-level and context-aware frameworks such as CORRECT~\cite{li2025everything}, CoderEval~\cite{yu2024codereval}, and HumanEvo~\cite{zheng2025humanevo} reveal integration failures that function-level benchmarks often miss, yet they are harder to standardize and compare across models. Manual and human-centered evaluations capture readability, reviewability, trust effects, and workflow realism, but they are expensive and difficult to scale~\cite{10606356, miah2024user, perry2023users}.

Taken together, the literature suggests that no single framework is sufficient for evaluating secure LLM-assisted coding. Static analysis is useful for breadth, dynamic testing is useful for behavioral evidence, repository-level benchmarks are necessary for contextual realism, and human-centered evaluation is necessary for deployment realism. The main gap in the current landscape is therefore not simply the lack of benchmarks, but the lack of an integrated evaluation pipeline that combines these complementary signals into a more trustworthy picture of security performance.

\section{Human Oversight and Human-Factor Risks}
\label{sec:human_role}
The integration of LLMs into coding tasks has significantly advanced software development, yet their inherent limitations necessitate human involvement to ensure code quality, correctness, and security~\cite{liu2024agents4plc, kathiresan2025human, tie2024llms, dou2024s}. Human oversight is often treated as a key safeguard, whether through direct review, feedback loops, or human-in-the-loop validation~\cite{li2026you, xia2023conversational, firouzi2026persistent, olaussonself, zheng2024opencodeinterpreter}. However, human participation is not uniformly protective: the same workflows that enable developers to detect flaws and clarify intent can also introduce vulnerabilities through over-trust, automation bias, weak prompt engineering, and insufficient review practices.

Accordingly, this section treats human-AI interaction as a dual-role factor in secure LLM-assisted coding. Human developers and reviewers remain critical for interpreting requirements, auditing outputs, and identifying flaws that automated systems miss, yet empirical evidence also shows that LLMs can reduce scrutiny, create false confidence, and distort responsibility in ways that degrade security \cite{hamer2024just, khoury2023secure, mohsin2024can, nazzal2024promsec}. To address \textbf{RQ3}, we therefore review the literature not only on human oversight as a mitigation mechanism, but also on human-AI collaboration as a source of security and accountability risk.

\subsection{\textbf{Limitation of Automated Feedback}}
A common approach to improving LLM-generated code is to design models capable of providing their own feedback and attempting self-correction, a process often referred to as self-repair or self-criticism~\cite{olaussonself, dou2024s}.
This typically involves the model generating an initial code snippet, executing it against test cases, and then using the resulting error messages or diagnostics as execution feedback to identify and rectify its own errors~\cite{zheng2024opencodeinterpreter, dai2025feedbackeval}. For instance, Agents4PLC~\cite{liu2024agents4plc} incorporates a dedicated debugging agent that provides fixing advice back to the coding agent when code fails, drawing inspiration from existing automated code repair methods. It can also leverage previously generated patches in a self-RAG context during the debugging phase to improve repair effectiveness. Similarly, frameworks like OpenCodeInterpreter~\cite{zheng2024opencodeinterpreter} involve prompting LLMs (e.g., GPT-4) to intentionally generate incorrect code, then using the resulting error messages as cues for correction. This process mimics a real-life debugging cycle. StepCoder~\cite{dou2024stepcoder} also uses reinforcement learning with compiler feedback to guide the exploration of the output space for LLMs, aiming to improve code generation quality.

Despite recent progress, models lack the formal correctness guarantees and specialized support required for industrial applications. In Programmable Logic Controller (PLC) programming, for example, they often struggle to satisfy the rigorous logical and safety constraints essential for industrial control systems~\cite{liu2024agents4plc}.
Moreover, LLMs have difficulty executing complex coding tasks and frequently produce suboptimal algorithms, resulting in issues like timeout errors~\cite{olaussonself, chidambaram2024socratic, dou2024s}. Even when directed to use efficient solutions like dynamic programming, LLMs can still create inefficient code~\cite{qiu2024efficient}.

Additionally, LLMs can misunderstand programming specifications, which leads to generated code that does not align with the user's original intent. When the generated code deviates significantly from the ground truth, the coarse-grained error messages produced during automated feedback cycles often prove insufficient for effective improvement~\cite{tian2025fixing}.

When LLMs are employed in coding assistance, they might inadvertently bring about new risks. For instance, feedback-driven efforts to rectify identified security flaws may result in the LLM inserting defects into previously flawless code~\cite{chong2024artificial}. Additionally, code produced by LLMs frequently omits defensive programming techniques, rendering it less secure and susceptible to vulnerabilities such as buffer and integer overflows~\cite{licorish2025comparing}. Furthermore, this machine-generated code tends to be more intricate per line compared to code crafted by humans, which might complicate subsequent maintenance and analysis~\cite{dou2024s}. The perceived advantages of LLMs can engender an unfounded sense of productivity and security, given that LLMs might create flawed or weak code and misleading recommendations. Excessive dependence on these models can result in time wastage and suboptimal tool utilization~\cite{tie2024llms}.

Alongside security issues, LLM-assisted coding tasks also face computational challenges. The cost of enabling self-repair might lead to only slight improvements, which are inconsistent across various tasks or datasets and can occasionally be negligible~\cite{olausson2023self}. This is mainly due to the limited capacity of the model to introspect and reliably generate effective feedback on its own mistakes~\cite{olausson2023self, chidambaram2024socratic}.

Finally, LLMs show decreased efficiency in processing feedback aimed at rectifying multiple errors at once, indicating that such feedback might be more effective if it is broken down into simpler steps to enhance model performance~\cite{chen2023improving}.

\subsection{\textbf{Incorporating Human Feedback}}
In addressing the limitations inherent in code generation by LLMs, human feedback is essential. This is commonly implemented through methods such as Reinforcement Learning from Human Feedback (RLHF) and Human-in-the-Loop (HITL) systems~\cite{ouyang2022training, kathiresan2025human, wong2024aligning}. Such strategies employ human insight, proficiency, and subtle comprehension to steer LLMs toward producing more accurate, secure, and coherent code. It's crucial that users verify their code's security and integrity~\cite{fu2024security, pearce2022asleep}, while developers should remain alert and exercise caution with generative AI~\cite{toth2024llms}. The reliability of LLM-generated code, including suggested improvements, is typically variable and necessitates human oversight to ensure consistency~\cite{elgedawy2024ocassionally, belozerov2025secure}. Developers are warned against placing complete trust in either AI-generated or AI-revised code fragments~\cite{hamer2024just}.

At the same time, incorporating human feedback does not automatically eliminate risk. The security benefits of RLHF, HITL, and related feedback-driven workflows depend heavily on reviewer expertise, the quality and consistency of the feedback provided, and the surrounding workflow design~\cite{belozerov2025secure, firouzi2026persistent, kathiresan2025human, li2026you, mohsin2024can, tie2024llms}. Poorly calibrated trust, superficial review, or misplaced confidence in AI-revised outputs can still allow insecure code to pass through development pipelines, while responsibility for failures may become blurred across model designers, tool providers, and human reviewers~\cite{belozerov2025secure, hamer2024just, khoury2023secure, perry2023users}. Human feedback should therefore be understood not as a guaranteed safeguard, but as a socio-technical control whose effectiveness depends on how the collaboration is structured and audited.

\noindent\textbf{Reinforcement Learning from Human Feedback (RLHF).} A prominent method for aligning model behaviour is RLHF, a technique that fine-tunes LLMs using human preferences as a reward signal~\cite{ouyang2022training, brown2020language, bai2022constitutional, wang2024reinforcement}. The standard pipeline involves supervised fine-tuning on human demonstrations, followed by the training of a reward model using human-ranked output comparisons. In the final stage, the model is refined via Proximal Policy Optimization (PPO)~\cite{ouyang2022training}, using the reward model as a signal to guide the model toward preferred responses.

RLHF has significantly improved the performance of models such as Codex and GPT across various tasks~\cite{wong2024aligning, ouyang2022training}. However, the method is constrained by the high cost and inconsistency of human feedback. Relying on individual evaluators introduces subjectivity and bias, which can be counterproductive in uncertain domains where input may be unreliable. Furthermore, the substantial labor required and the limited scale of human-labeled datasets often restrict the efficiency and measurable impact of this alignment strategy.

\noindent\textbf{Crowd-Sourced RLHF (cRLHF).} To mitigate the inherent biases and constraints associated with individual evaluators, Wong et al.~\cite{wong2024aligning} introduced cRLHF, a framework that mitigates the biases of individual evaluators by aggregating feedback from multiple annotators. Using Bayesian inference, the system combines multi-source assessments—including line-by-line accuracy—into a "coordinated score" between 0 and 1. This score serves as a direct reward for model refinement, simplifying the reward modeling process while ensuring outputs align with diverse human expectations. Empirical results demonstrate that this collaborative approach effectively reduces model bias and enhances the quality of generated code

Crowdsourced helps reduce bias and improve evaluation objectivity by aggregating diverse human inputs. However, its effectiveness is often constrained by small sample sizes and the logistical challenge of recruiting expert annotators for large-scale studies \cite{wong2024aligning}.

\noindent\textbf{Human-in-the-Loop (HITL)} Another crucial approach is HITL testing, which integrates human oversight at critical junctures to supplement automated evaluations~\cite{kathiresan2025human}. This approach addresses the non-deterministic nature of model outputs by using automated tools to flag high-impact or low-confidence results for expert review. Human feedback on the accuracy and safety of these outputs is then used to refine prompts and triage criteria. This iterative process, which may include RLHF for response optimization~\cite{ouyang2022training}, leverages models to generate adversarial test cases and summarize large datasets, ultimately improving both the reliability and precision of the system.

HITL improves the reliability and safety of model-integrated systems, especially in high-stakes domains, by combining large-scale automated analysis with targeted human judgment~\cite{kathiresan2025human}. Human reviewers can catch nuanced errors, biases, and contextual issues that automated tools often miss, although the approach remains selective and is limited by the scale of expert feedback available.

\noindent\textbf{Conversational Automated Program Repair (Conversational APR).} Conversational APR iteratively generates patches and incorporates validation feedback, such as test results, outputs, and error messages, to refine subsequent repairs~\cite{xia2023conversational}. By combining prior patches with feedback in each round, it helps LLMs avoid repeating mistakes and better understand program semantics. Successful patches can also be varied through additional conversational turns.

This approach can substantially improve repair quality by leveraging natural-language hints and human clarification; indeed, a brief human correction may be more effective than greatly increasing the number of generated samples. However, the method can be computationally expensive because it often requires many LLM interactions~\cite{tian2025fixing}.

\noindent\textbf{Conversational Semi-Automatic Repair Framework (Cref).} This system employs LLMs specifically designed to assist programming tutor~\cite{yang2024cref}. It employs interactive processes involving tutor insights, solution narratives, and test cases that fail. Cref handles everything in one dialogue session and advances through specific stages: initially supplying LLMs with programming problems, erroneous code, and tutor assistance; if the code is still erroneous, it revisits all prior data and solution outlines; and if further correction is needed, it utilizes historic entries alongside failing test cases of the most recent code fix. This methodology notably enhances code generation efficiency via natural language prompts. Conversational APR~\cite{xia2023conversational} strategically circumvents the replication of past errors in patches and effectively uses previously overlooked test case details. Cref has achieved substantial progress in Average-5 scores for code repair, cutting debugging times by 71.2\% and costs by 69.9\% in real-world corporate scenarios. It also boosts tutoring success and enriches students' learning experiences. The achievements of Cref indicate its potential for wider use cases, such as automating code review processes by refining code based on reviewer feedback.

\noindent\textbf{Socratic Human Feedback (SoHF).} SoHF~\cite{chidambaram2024socratic} is a form of expert feedback in which programmers guide LLMs through iterative natural-language prompts, much like a tutoring session. Using strategies such as test reiteration, error highlighting, and requirement clarification, experts provide step-by-step guidance across multiple debugging rounds. This approach can substantially improve performance: with expert Socratic guidance, LLMs were able to solve up to 74\% of problems they initially failed. However, existing studies focus mainly on expert programmers, and some do not account for the time required for debugging, leaving its broader applicability unclear.

\noindent\textbf{Imitation Learning from Language Feedback (ILF).} Chen et al.~\cite{chen2023improving} introduced ILF, a framework that incorporates natural language feedback during the training phase rather than solely at inference. The process begins by identifying code that fails unit tests and obtaining human-written feedback for those errors. A specialized refinement model, $\pi_{Refine}$, uses this feedback to generate corrected versions of the code, which are then used to fine-tune the primary model. This iterative approach, which effectively minimizes the KL divergence from the target distribution, requires minimal human-authored feedback and remains independent of such feedback during testing. Evaluation on the MBPP benchmark showed that ILF increased the pass@1 rate of a 6.1B parameter model by 38\% relatively and 10\% absolute. Despite these gains, the framework is limited by the high cost of human feedback and the difficulty $\pi_{Refine}$ faces when addressing multiple, simultaneous errors.

Expert review is vital for auditing and refining code produced by automated systems to ensure functional correctness and security \cite{dai2025feedbackeval, liu2024agents4plc, wong2024aligning}. Professional developers are uniquely equipped to implement defensive programming and guardrails that automated tools frequently overlook, particularly in security-critical environments \cite{chong2024artificial}. By leveraging deep domain knowledge and context awareness, human reviewers can resolve ambiguous requirements and adhere to industrial standards more reliably than current computational models \cite{licorish2025comparing}. Despite being resource-intensive and difficult to scale, manual oversight remains a necessary safeguard against the structural complexity and subtle vulnerabilities common in machine-authored software \cite{zhou2020learning, chong2024artificial}.

Overall, human expertise continues to be crucial for addressing the shortcomings of LLMs, particularly in handling intricate, security-sensitive, or unclear tasks where LLMs face challenges related to efficiency, ambiguous requirements, and the creation of robust, secure, and maintainable code~\cite{chong2024artificial, licorish2025comparing, yang2024cref}. It is most effective to regard LLMs as an enhancement to human programmers rather than a replacement. The cooperation between human skills and LLM capabilities, especially through iterative and focused feedback, is essential for the progress of AI-supported programming~\cite{licorish2025comparing}.

Taken together, the literature suggests that human oversight is both indispensable and fragile in LLM-assisted coding. Human expertise remains essential for clarifying intent, evaluating security-sensitive outputs, and correcting model failures, but the introduction of LLMs can also distort developer behavior through over-reliance, reduced scrutiny, and miscalibrated trust \cite{belozerov2025secure, hamer2024just, khoury2023secure, mohsin2024can, perry2023users, tihanyi2024neutral}. The practical challenge is therefore not simply to keep a human in the loop, but to design human-AI workflows in which oversight is informed, accountable, and appropriately matched to the risks of the task.

\section{Open Challenges and Research Directions}
\label{sec:future}

The reviewed literature points to three persistent challenges for secure LLM-assisted coding: a mismatch between probabilistic generation and deterministic security requirements, unreliable or incomplete evaluation pipelines, and fragile human-AI collaboration shaped by over-trust, weak prompting, and limited accountability. This section synthesizes these challenges into a research and deployment agenda, then outlines governance, practical recommendations, and broader implications for academia and industry.

\subsection{Core Challenges}
Our study indicates that the security concerns associated with LLM-assisted coding do not merely consist of isolated bugs but rather merge into three interrelated, major meta-challenges. These three vulnerabilities form the core of the issue and require a comprehensive approach for resolution.

\subsubsection{Probabilistic Models vs. Deterministic Security}
The core issue is a fundamental discrepancy between the function of LLMs and the demands of secure software development. LLMs operate as probabilistic, pattern-driven models designed to predict the next probable token based on the statistical trends found in their training data. They do not possess a profound, semantic comprehension of security principles, threat vectors, or the intricate context that differentiates secure from insecure code. Their key aim is functional plausibility and fluency in language, which means they aim to generate working code without an inherent focus on security vulnerabilities. 

This fundamental limitation manifests through several significant issues. Firstly, LLMs are trained on extensive publicly available code, such as the open-source collections on GitHub, which tend to contain vulnerabilities, outdated dependencies, and insecure coding standards. As a result, the models learn to emulate these issues. Research repeatedly shows that a significant proportion of code produced by LLMs is inherently insecure; for example, studies reveal that around 40\% of snippets from models such as GitHub Copilot contain vulnerabilities. This happens because the model's objective is to replicate patterns, and insecure patterns are abundant in the training data. Secondly, when LLMs are used for APR and patching vulnerabilities, they may produce corrections that seem functionally sound for a limited set of tests but remain vulnerable to more advanced or unforeseen threats. The patch may fix the current issue but inadvertently create a new, subtler vulnerability or only partially mitigate the risk, leaving it exploitable. Third, LLMs often suggest or use deprecated libraries, outdated APIs, or cryptographically vulnerable algorithms because such insecure patterns remain within their training sets and are suggested without recognising their compromised state. 

Although the common explanation often blames bad training data, a more detailed analysis uncovers a deeper architectural flaw. The central problem is a semantic gap between statistical association and logical deduction. LLMs, as designed, do not reason about security attributes; they determine statistical probabilities. Even with security-focused prompts, models struggle because they are not equipped for abstract reasoning about possible threat vectors or security invariants. This indicates that merely cleaning the training data, while necessary, will not fully solve the problem. It is similar to teaching someone to write grammatically by exposing them to a large library without explicitly teaching grammatical rules. They might mimic style and common expressions, but cannot reason from grammatical fundamentals. Thus, to achieve genuinely secure code generation requires more than just data refinement; it points to a necessity for profound changes in model design, training goals, or the incorporation of symbolic reasoning abilities.

\subsubsection{The Unreliability of Ground Truth}
The second significant obstacle is the inadequacy of current methods for assessing the security of code generated by LLMs. This leads to an evaluation crisis, where benchmark scores can be deceptive, potentially inflating a model's genuine security capabilities and creating an illusion of progress. Without reliable measurement, improvements cannot be accurately made. This crisis stems from various shortcomings in current evaluation practices. Firstly, many primary benchmarks for code generation, such as HumanEval~\cite{chen2021evaluating}, focus on functional correctness for small, self-contained programming tasks. However, these benchmarks often have a limited range and size, insufficient for detecting subtle logical mistakes or complex security weaknesses, resulting in exaggerated performance metrics that don't represent true robustness. Besides, the security evaluation of code heavily relies on SAST tools like CodeQL~\cite{codeql} and Bandit~\cite{bandit}, which are known to be neither sound nor comprehensive, displaying considerable rates of false positives and false negatives. Moreover, these tools typically assess code in isolation, lacking the broader repository context (like data flows across multiple files) necessary to uncover many intricate real-world vulnerabilities. Although dynamic analysis can surpass static analysis in certain areas, it is inherently limited by the quality and breadth of the test cases employed. Most benchmarks fail to mimic the complexity of real-world software development, where newly created code must be correctly and securely integrated into a large, existing, and complex repository. 

There is a clear progression in evaluation paradigms, shifting the primary question from "Does the code work?" to "Can the code be exploited?". Initial benchmarks like HumanEval~\cite{chen2021evaluating} and MBPP~\cite{austin2021program} concentrated on functional correctness, utilising a cooperative set of predefined inputs. In contrast, a new wave of benchmarks is surfacing that adopts a more adversarial approach. Frameworks such as SecRepoBench~\cite{dilgren2025secrepobench}, CodeLMSec~\cite{hajipour2024codelmsec}, and BAXBENCH~\cite{vero2025baxbench} are crafted to test vulnerabilities under more authentic conditions. They focus on repository-level context, dynamic execution of exploits, and resilience against adversarial inputs. This transition is significant as it redefines a good model not merely as one that passes tests but as one that withstands attempts at compromise. The future of effective LLM security evaluation lies in advancing and utilising these more demanding, adversarial, and context-sensitive benchmarking frameworks.

\subsubsection{Human-AI Trust-Vulnerability}
The third significant challenge emerges at the complex intersection where human developers engage with AI assistants. While human oversight is generally considered the most reliable defence against the weaknesses of LLMs, it also presents significant vulnerabilities. This forms a trust-vulnerability nexus, where cognitive biases, different expertise levels, and misaligned trust can compromise the security of the entire development process. This connection presents itself in several concerning ways. First, a dangerous bias has surfaced where developers, especially those using AI coding assistants, generate code with lower security standards, even while being overly confident about its safety. This situation generates a misguided sense of security, which diminishes thorough examination and leads to the premature integration of flawed or insecure code into operational systems. Furthermore, the security of output from LLMs is largely dependent on the user's skill in creating precise, security-conscious prompts and critically assessing the generated code. Inexperienced programmers or developers without specific security training are especially likely to create vague prompts and accept superficially suitable but insecure code, thereby adding new vulnerabilities to the software supply chain. Additionally, in the realm of software engineering, the notion of trust in LLMs is currently vague, inconsistently evaluated, and poorly comprehended. This leads to a precarious miscalibration, where a developer's trust level does not match the model's actual effectiveness for a task. This could lead to either over-trust, resulting in uncritical acceptance of flawed code, or under-trust, inhibiting the use and advantages of the technology.

The common advice of simply training developers more is necessary, but not enough to tackle this challenge. While it is crucial to educate on security, the main issue is not merely a lack of knowledge but one of human-computer interaction (HCI). The essential challenge is around calibrating trust. Developers need efficient ways to discern when and how much to trust the LLM's output, as well as what aspects require closer examination. This shifts the focus from merely training to navigating a complex design challenge. Future research and tool development should aim to build Human-AI collaboration systems that actively aid in this calibration. Potential solutions include AI systems that communicate their uncertainty, provide verifiable evidence for their code suggestions, or automatically flag high-risk code segments that demand mandatory, thorough human assessment. This approach seeks to shift the human role from a passive, potentially susceptible reviewer to an empowered, critical, and effective guardian of code quality and security.

\subsection{A Research Roadmap for Trustworthy LLM-Assisted Coding}
Tackling the triad of challenges presented earlier demands a collaborative, cross-disciplinary research approach. We organize this roadmap around three core pillars: constructing inherently secure models, developing more reliable and realistic security benchmarks, and improving the calibration of human-AI collaboration. A fourth, cross-cutting requirement (accountability, auditability, and governance) is discussed separately in Section~\ref{subsec:aag} because it conditions how progress in all three pillars can be translated into trustworthy real-world deployment.

\subsubsection{Towards Inherently Secure Models}
The key aim is to evolve from a reactive, post-incident security stance towards developing LLMs that are inherently secure. Achieving this goal involves integrating security principles into the model's fundamental architecture and training frameworks, rather than considering security as an afterthought. Security-aware architectural design and training are crucial for upcoming research, directing focus towards innovative model structures and training goals that surpass mere next-token forecasting. The prevailing approach, which chiefly values functional accuracy and fluency, falls short for applications critical to security. Future inquiries should explore incorporating security-driven loss functions that specifically discourage code creation with established vulnerability patterns (CWEs). Another promising path is crafting hybrid architectures that merge the pattern-detection capabilities of neural networks with symbolic reasoning elements adept at grasping and imposing security properties and constants. 

Advanced Security-Oriented Fine-Tuning provides a practical solution, given the formidable costs associated with initiating foundation model pre-training. Techniques for Parameter-Efficient Fine-Tuning (PEFT), like Low-Rank Adaptation (LoRA), have proven effective for tailoring pre-trained models to secure code generation without the extensive computational demands of full retraining. A pivotal area of research focuses on developing and curating extensive, high-quality datasets featuring confirmed vulnerability corrections to instil secure coding norms in models. Additional study is warranted to comprehend how varying granularities of data—such as file-level, function-level, or detailed block-level modifications—affect the security of the refined model's outputs. A notable concern is safety degradation, where fine-tuning a model even with benign, task-specific data may unintentionally jeopardize its overall safety alignments. This underlines the necessity for Safety-Aware Fine-Tuning (SAFT) frameworks designed to automatically identify and exclude potentially harmful or vulnerability-inducing data from the refinement set or to guide gradient updates away from hazardous parameter territories via safety-aware probes. 

Formal Methods for Verifiable Security and Correctness represent the most thorough approach to ensuring software security through formal validation. Although directly producing formally verified code via LLMs is currently unsuitable except for the simplest programs, a highly promising area merges LLMs with formal methods tools. In this context, the LLM serves as a verification assistant rather than the definitive code author. For instance, frameworks like Marmaragan~\cite{cramer2025verifying} utilise an LLM to produce formal annotations, such as pre- and post-conditions, loop invariants, and contracts, for existing code in languages like SPARK. These annotations are then subjected to checks by a dedicated formal verifier. The LEMUR calculus lays down a theoretical base for this interaction, conceptualising a coherent loop where the LLM suggests properties and an automated reasoner seeks to validate them. Similarly, LLMLift employs an LLM to generate program invariants necessary for proving the equivalence of code translated between languages. This hybrid approach, blending verifier-in-the-loop, significantly harnesses the pattern-recognition and creative-generation capacities of LLMs while rooting them in the mathematical exactness of formal verification, paving the way for constructing more reliable systems.

\subsubsection{Next Generation of Security Benchmarks}
To advance and monitor progress effectively within the community, it is essential to develop a new generation of evaluation benchmarks that are more trustworthy, realistic, and thorough than those currently widespread. 

\noindent\textbf{Holistic, Contextual, and Repository-Level Evaluation:} The field must move decisively beyond evaluating isolated, function-level code snippets created in isolation. Future benchmarks should be designed at the repository level to be impactful. This requires furnishing the LLM with the relevant context—such as dependency structures, data-flow across files, and existing API definitions—needed to produce code that integrates accurately and securely into a larger, pre-existing codebase, not just correct in isolation. SecRepoBench~\cite{dilgren2025secrepobench} is an innovative initiative in this aspect, formulating code completion tasks from actual vulnerability patches in C/C++ projects and employing the projects' own unit test suites for validation. The HumanEvo~\cite{zheng2025humanevo} benchmark enhances this by offering a historically accurate repository context, restoring a project to its state just before a target function was initially developed, thereby creating a more realistic evaluation scenario. 

\noindent\textbf{Dynamic and Exploit-Driven Security Assessment:} Sole reliance on static analysis to identify vulnerabilities is inadequate due to its inherent lack of precision. Future security evaluations must be dynamic, monitoring code during execution, and, where feasible, driven by the execution of real exploits. 

\noindent\textbf{Benchmarking the Human-AI System:} A crucial step for future evaluation research is to transition from assessing the AI alone to benchmarking the performance of the entire human-AI system. This would entail devising evaluations that mimic a real-world developer workflow. Metrics could encompass the security of code generated by a developer with an AI assistant, the time and effort needed for a human reviewer to identify and amend AI-induced flaws, or the efficiency of AI-aided review tools in enhancing a developer's security performance. This comprehensive approach would deliver a more accurate depiction of these tools' impact on software security.

\subsubsection{Calibrating Human-AI Collaboration}
Given that human oversight will remain an essential component of secure software development for the foreseeable future, research must focus on making this collaborative process as effective and secure as possible. The goal is to transform the human from a potential point of failure into an empowered and critical partner.

\noindent\textbf{Modelling and Measuring Trust to Mitigate Over-reliance:} The well-documented risk of automation bias and developer over-trust is a primary human-related threat in LLM-assisted coding. To counter this, research drawing heavily from the field of Human-Computer Interaction (HCI) is needed to formally define, model, and measure the concept of trust in the specific context of software engineering. The objective is not simply to maximise trust, but to develop systems that can actively calibrate a user's trust to align with the model's actual capabilities and reliability for a given task. This could be achieved through novel user interfaces that visualise a model's uncertainty about its own output, provide verifiable evidence for its suggestions, or automatically flag high-risk or low-confidence code segments that require mandatory, heightened human scrutiny.

\noindent\textbf{Developing Tools for Secure Prompting and AI-Assisted Review:} The quality of the interaction between the developer and the AI is paramount. Research should focus on creating tools and interaction paradigms that facilitate more secure and effective collaboration. Promising directions include:
\begin{itemize}
    \item Interactive and Conversational Repair: Frameworks that use a dialogue-based process of iterative feedback and refinement have shown significant promise. Systems like Cref~\cite{yang2024cref} and conversational APR~\cite{xia2023conversational} allow a developer to guide the LLM toward a correct and secure solution through a natural conversation, providing error messages or clarifications that help the model understand the program's semantics and avoid repeating mistakes.
    \item Guided Reasoning and Socratic Feedback: Techniques like Socratic Human Feedback (SoHF)~\cite{chidambaram2024socratic}, where an expert programmer guides the LLM with a series of probing questions and step-by-step debugging hints, can enable models to solve complex problems that they would fail at on their own. This approach turns the interaction into a collaborative problem-solving session rather than a simple request-response transaction.
    \item AI-Powered Review Tools: The development of tools that integrate LLM-based vulnerability analysis directly into the IDE can augment the developer's abilities. Tools like CODEGUARDIAN~\cite{dozono2024large} have been shown to make developers both faster and more accurate at identifying security flaws in code.
\end{itemize}

\subsection{Accountability, Auditability, and Governance}
\label{subsec:aag}
A recurring theme across the literature is that the risks of LLM-assisted coding cannot be reduced to model quality alone~\cite{treude2026accountable, taldenko2026human, chow2025large, chow2026dialogue, kathiresan2025human, li2026you, NIST_AI_RMF_2023}. Even if a model is improved through better data curation, safer decoding, or stronger evaluation pipelines, real-world deployment still raises unresolved questions of responsibility, traceability, and governance~\cite{chong2024artificial, chow2025large, kathiresan2025human}. These issues become especially acute when insecure code is generated collaboratively by humans and AI systems or when agentic systems perform multi-step actions with limited human intervention~\cite{chow2025large, li2026you}. Accordingly, accountability, auditability, and governance should be treated not as peripheral policy concerns, but as cross-cutting conditions for the trustworthy deployment of LLM-assisted coding systems.

\noindent\textbf{Responsibility for insecure outputs.}
A fundamental challenge concerns responsibility allocation when insecure or harmful code is introduced through AI-assisted development~\cite{chong2024artificial, chow2025large, chow2026dialogue, li2026you}. In practice, responsibility may be distributed across several actors: the model developer, the provider of the coding assistant, the organization deploying the tool, the engineer who accepted the output, and the reviewer who approved it. This diffusion of responsibility can create accountability gaps, especially when AI-generated code is treated merely as a productivity aid rather than as an active decision-shaping component of the development process~\cite{chow2025large, chow2026dialogue, ma2024code}. For secure deployment, organizations must therefore adopt explicit policies stating that AI assistance does not eliminate human accountability for the final artifact.

\noindent\textbf{Auditability and traceability.}
Secure use of LLM-assisted coding also requires the ability to reconstruct how a code artifact was produced~\cite{NIST_AI_RISK_2024,chow2026dialogue, li2026you, treude2026accountable}. This includes preserving relevant prompts, retrieved context, intermediate model outputs, tool invocations, validation results, and subsequent human edits. Without such traceability, it becomes difficult to investigate incidents, reproduce failures, or determine whether a vulnerability emerged from the model, the prompt, the retrieval source, the toolchain, or the human review process~\cite{chow2026dialogue,li2026you}. Auditability is thus not only a compliance concern; it is also a practical prerequisite for debugging, incident response, and continuous security improvement.

\noindent\textbf{Liability in regulated and high-stakes environments.}
The accountability problem becomes more pressing in regulated domains or safety-critical contexts, where software faults may trigger legal, financial, or societal harm~\cite{euai,chow2025large, chow2026dialogue}. In such settings, organizations cannot rely on opaque AI-generated artifacts without rigorous documentation, testing, and approval procedures. The introduction of LLM-assisted coding into domains such as healthcare, finance, public infrastructure, or industrial control raises difficult questions about due diligence, certification, and evidentiary standards for demonstrating that adequate safeguards were applied.

\noindent\textbf{Governance as a socio-technical control layer.}
These concerns suggest that governance must be treated as a first-class layer of defence rather than as an afterthought \cite{euai,chong2024artificial, chow2025large, chow2026dialogue}. Effective governance includes policies for prompt and data handling, restrictions on the use of external tools and proprietary context, model and plugin vetting, mandatory review requirements for security-critical code, and documentation of human oversight and approval pathways. In the case of agentic systems, governance should additionally include action boundaries, escalation rules, logging of tool-mediated behavior, and explicit mechanisms for human interruption and override \cite{chow2026dialogue,li2026you}. Overall, accountability, auditability, and governance connect technical security properties with real deployment conditions and help ensure that human-AI collaboration remains reviewable, contestable, and aligned with institutional responsibility.

Overall, accountability, auditability, and governance should be understood as integral components of secure LLM-assisted coding. They connect technical security properties with real deployment conditions and help ensure that human-AI collaboration remains reviewable, contestable, and aligned with institutional responsibility.

\subsection{Practical Insights and Actionable Recommendations}
Organisations and software development teams have the opportunity to promptly implement specific measures to reduce security risks linked with incorporating LLMs into their processes. Here are seven practical recommendations that offer a comprehensive defence strategy for both leveraging the advantages of LLMs and addressing their potential risks.

\noindent\textbf{Adopt a Zero-Trust Policy for AI-Generated Code.} Treating all code generated by an LLM as untrusted third-party input is the most indispensable rule, no matter how advanced the model or how specific the prompt. Such code should undergo stringent security assessments, peer review, and extensive testing akin to what is applied to code from external or junior sources. It is imperative that LLM-generated code never be deployed directly to production without exhaustive human supervision and a series of automated security evaluations.

\noindent\textbf{Invest in Continuous Developer Training for Secure AI Interaction.} Human developers serve as the ultimate safeguard, with their knowledge being a crucial resource. Organisations need to prioritise ongoing training initiatives to enhance developers' understanding of the inherent limitations and particular security challenges associated with LLMs. This educational effort should go beyond standard secure coding practices to encompass AI-centric skills, including secure prompt engineering (e.g., crafting clear, precise, and security-conscious prompts), understanding common vulnerabilities linked to LLMs, and methods for critically assessing AI-generated code, including identifying indicators of model overconfidence or factual inaccuracies.

\noindent\textbf{Integrate Multi-Layered, Automated Security Analysis into the CI/CD Pipeline.} Utilizing only one security tool or method is inadequate for addressing the variety of possible issues in code generated by AI. An effective approach necessitates establishing a pipeline that integrates multiple levels of automated analysis that complement each other. This should encompass: Static Application Security Testing (SAST) to detect frequent vulnerability patterns within the source code prior to execution; Dynamic Application Security Testing (DAST) to discover runtime errors and vulnerabilities in the operational application; and Software Composition Analysis (SCA) to pinpoint and control vulnerable or outdated third-party libraries and dependencies, as LLMs are often known to recommend such components.

\noindent\textbf{Mandate Rigorous, Context-Aware Human Review for Security-Critical Code.} Although all AI-produced code necessitates human oversight, code that is to be implemented in security-sensitive functions must undergo a rigorous and obligatory review process. This encompasses code involved in authentication, authorisation, encryption activities, session management, and input verification. Such evaluations should be performed by senior developers or specialised security professionals capable of analysing the code not only independently, but also within the broader context of the application, taking into account its interactions with various components and data pathways.

\noindent\textbf{Develop and Enforce Strict Organisational Policies for Prompt Engineering and Data Handling.} Organisations need to create and implement well-defined governance policies outlining how developers should engage with LLMs, especially those from third parties. Such policies should clearly forbid embedding sensitive or proprietary data—like API keys, internal algorithms, trade secrets, or Personally Identifiable Information (PII)—into prompts given to external models. In the case of applications utilising Retrieval-Augmented Generation (RAG), these policies must guarantee that internal knowledge bases used for retrieval are securely protected, access-controlled, and sanitised to avoid unintended data breaches and prevent malicious RAG poisoning attacks.

\noindent\textbf{Prioritise and Vet the AI/ML Component Supply Chain.} Before introducing a new tool, plugin, or open-source model based on LLMs, it is essential for organisations to perform comprehensive security due diligence. This process includes assessing the vendor's security practices, data privacy and retention policies, and model training procedures.5 In the case of open-source models obtained from public repositories, a standardised intake process should be adopted. This process should involve initial red-teaming to explore adversarial vulnerabilities and bias evaluations to highlight any potential ethical risks prior to integrating the model into internal development processes.

\noindent\textbf{Establish Clear Lines of Accountability for AI-Assisted Development.} The legal and regulatory framework concerning liability for mistakes and weaknesses in AI-generated code is unclear and continues to develop. In such a scenario, organisations must implement well-defined internal policies that delineate accountability. The human developer and engineering team should bear the final responsibility for the security, accuracy, and quality of the completed software, rather than the AI tool utilised. Keeping comprehensive documentation of the human oversight, amendment, and approval processes for all AI-generated code is vital for showing thoroughness and mitigating legal liabilities.

\subsection{Implications for Academia and Industry}
The rise of LLM-assisted programming brings about profound implications extending beyond technical challenges alone. It is reshaping research directions in academia and demanding new strategic methods for risk management and governance within the industrial sector. Navigating this new era effectively requires a collaborative effort between these communities to balance the pursuit of innovation with the crucial requirement for security and reliability.

\subsubsection{Implications for Academia}
The issues highlighted by this survey indicate the necessity for a pivotal change in academic approaches to software engineering and artificial intelligence research. 

Traditionally, success for code-generating LLMs has been measured by their performance on tests like HumanEval~\cite{chen2021evaluating}, which largely focus on solving algorithmic puzzles in controlled environments. It is critical for the academic field to shift its focus toward emphasising security as a primary research goal. This includes fostering research on model architectures designed with security in mind, developing robust defences against adversarial threats, and creating realistic evaluation benchmarks that emphasise context-rich security scenarios. Adopting this new focus may require accepting compromises, such as models that are more secure but show reduced performance on specific functional tasks being deemed more valuable scientific contributions. 

The challenge of ensuring the security of LLM-assisted coding is not one that can be tackled by computer science in isolation. For instance, the Human-AI Trust-Vulnerability Nexus is inherently a Human-Computer Interaction (HCI) issue, demanding insights from psychology and cognitive science to effectively address and mitigate biases. Additionally, tackling concerns related to legal liability, data privacy, intellectual property, and algorithmic fairness requires collaboration with experts in law, ethics, and social sciences. The most significant future research efforts in this field will be fundamentally interdisciplinary, blending technical solutions with a detailed understanding of their human, societal, and legal implications to develop systems that are not only technically sound but also genuinely trustworthy. 

Besides the practical aspect of constructing and testing more secure systems, there is an urgent need for foundational research aimed at understanding why these systems fail. This involves interpretability studies using techniques like representation engineering to explore how LLMs internally comprehend security concepts such as vulnerability or sanitisation. It also includes theoretical research seeking to outline the connection between a model's structural features, its training data, and the security assurances of its outputs. Gaining a deeper, foundational understanding of these systems is essential for advancing beyond empirical trial-and-error toward the systematic design of verifiably secure AI.

\subsubsection{Implications for Industry}
For industries, the primary challenge is to strategically manage the inherent conflict between leveraging the substantial productivity gains offered by LLMs and addressing the considerable security, legal, and reputational risks they pose. 

LLMs are fundamentally redefining the economic dynamics of cybersecurity. For attackers, LLMs commoditise intelligence, significantly reducing the cost to uncover vulnerabilities and engineer custom, long-tail assaults on a broad spectrum of less common software, rendering previously unattractive targets financially worthwhile. This generates a compelling market-driven incentive for organisations to bolster their AI security measures. On the flip side, there is a rising positive economic incentive. As the market for specialised LLM agents and tools evolves, aspects like security, reliability, and trustworthiness will become vital competitive advantages. Vendors capable of proving that their products deliver enhanced security and produce safer code will be able to demand a premium, encouraging a market that benefits those investing in secure AI advancements. 

The global regulatory framework for AI is quickly transitioning from broad ethical guidelines to specific, enforceable legal standards. Regulations such as the EU AI Act, which classifies AI systems based on risk, the NIST AI Risk Management Framework, and new standards like ISO 42001, are setting a new compliance landscape for organisations deploying AI~\cite{euai}. These rules will necessitate organisations to develop strong governance, risk management, and compliance strategies for their AI systems. Moreover, the ongoing uncertainty regarding legal accountability for damages created by flaws in AI-produced code will push companies to adopt stringent internal validation, oversight, and documentation procedures to display due diligence and reduce legal liabilities. In the complex and dynamic environment, industry cannot afford to await perfectly secure, one-size-fits-all AI models. The sole feasible approach is a proactive, defence-in-depth strategy that combines technology, processes, and people. This entails investing in a comprehensive security technology stack, forming robust internal AI governance committees with well-defined accountability, and crucially, cultivating a culture of security awareness and critical thinking among developers. In the end, as AI increasingly automates coding procedures, the uniquely human abilities to judge, understand context, and provide critical oversight become some of the most crucial and irreplaceable security assets an organisation holds.

\section{Conclusion}
\label{sec:conclusion}
The integration of LLMs into software development offers substantial productivity benefits but introduces systemic security risks. These vulnerabilities originate from a fundamental triad: the models' lack of semantic security awareness, the inadequacy of current benchmarks, and the cognitive biases inherent in human-AI interaction. Our review identifies risks ranging from insecure code generation to the propagation of outdated dependencies, highlighting that evaluation must move beyond functional correctness toward adversarial, context-specific security assessments. Furthermore, developer oversight remains a critical but fragile defense, requiring tools that support rather than mislead the user. To address these challenges, we propose a research roadmap focused on security-driven training, more stringent benchmarks, and robust collaboration frameworks. For practitioners, we advocate a defense-in-depth strategy rooted in a zero-trust approach to model-generated code. Ultimately, the secure adoption of this technology requires a coordinated effort between academia and industry to prioritize security-focused governance alongside innovation.

\bibliographystyle{ACM-Reference-Format}
\bibliography{ref}
\end{document}